\newcommand{\aap}{A\&A}

\newcommand{\apjl}{ApJL}
\newcommand{\apjs}{ApJS}

\documentclass[useAMS,usenatbib]{mn2e}
\usepackage{txfonts}

\usepackage{graphicx}
\usepackage{longtable}

\title[Herschel spectra of the red-supergiant VY CMa]
  {{\it Herschel} SPIRE and PACS observations of the red 
supergiant VY~CMa: analysis of the molecular line spectra
\thanks{{\it Herschel} is an ESA space observatory with science instruments provided by 
European-led Principal Investigator consortia and with important participation from NASA.}}
\author[Matsuura et al.]
  {Mikako Matsuura$^1$, J.\,A. Yates$^1$, 
   M.\,J. Barlow$^1$, 
   B.\,M. Swinyard$^{1,2}$,
   P. Royer$^{3}$, 
   J. Cernicharo$^4$, 
            \newauthor 
   L. Decin$^3$, 
   R. Wesson$^{1,5}$,  
   E.T. Polehampton$^{2, 6}$,
   J.A.D.L. Blommaert$^3$, 
   M.A.T. Groenewegen$^7$,
                \newauthor 
   G.C. Van de Steene$^7$
   and P.A.M. van Hoof$^7$ \\
  $^1$ Department of Physics and Astronomy, University College London, Gower Street, London WC1E 6BT, UK\\
  $^2$ Space Science and Technology Department, Rutherford Appleton Laboratory, Didcot, Oxfordshire, OX11 0QX, UK \\
  $^3$ Instituut voor Sterrenkunde, Katholieke Universiteit Leuven, Celestijnenlaan 200D, 3001 Leuven, Belgium \\
  $^4$ Laboratory of Molecular Astrophysics, Department of Astrophysics, CAB, INTA-CSIC, Ctra de Ajalvir, 
       km 4, 28850 Torrej\'on de Ardoz, Madrid, Spain \\
  $^5$ European Southern Observatory, Alonso de Cordova 3107, Casilla 19001, Santiago, Chile \\
  $^6$ Institute for Space Imaging Science, Department of Physics \& Astronomy, University of Lethbridge, Lethbridge, AB T1K 3M4, Canada \\
  $^7$ Royal Observatory of Belgium, Ringlaan 3, B-1180 Brussels, Belgium \\
  }
\date{Released 2013 Xxxxx XX}

\pagerange{\pageref{firstpage}--\pageref{lastpage}} \pubyear{2013}

\def\LaTeX{L\kern-.36em\raise.3ex\hbox{a}\kern-.15em
    T\kern-.1667em\lower.7ex\hbox{E}\kern-.125emX}

\begin{document}

\label{firstpage}

\maketitle

\begin{abstract}

We present an analysis of the far-infrared and submillimetre molecular 
emission line spectrum of the luminous M-supergiant VY~CMa, 
observed with the SPIRE and PACS spectrometers aboard the {\em Herschel 
Space Observatory}. Over 260 emission lines were detected in the 
190--650-$\mu$m SPIRE FTS spectra, with one-third of the observed 
lines being attributable to H$_2$O. Other detected species include CO, 
$^{13}$CO, H$_2^{18}$O, SiO, HCN, SO, SO$_2$, CS, H$_2$S, and NH$_3$.  
Our model fits to the observed 
$^{12}$CO and $^{13}$CO line intensities yield a $^{12}$C/$^{13}$C ratio 
of 5.6$\pm$1.8, consistent with measurements of this ratio for other 
M~supergiants, but significantly lower than previously estimated for 
VY\,CMa from observations of lower-J lines. 
The spectral line energy distribution for twenty SiO rotational lines 
shows two temperature components: a hot component 
at $\sim$1000\,K, which we attribute to the stellar atmosphere and inner 
wind, plus a cooler $\sim$200\,K component, which we attribute to 
an origin in the outer circumstellar envelope. 
We fit the line fluxes of $^{12}$CO, $^{13}$CO, H$_2$O and SiO,
using the  {\sc smmol}  non-LTE line transfer code, with a mass-loss rate of $1.85\times10^{-4}$\,$M_{\odot}$\,yr$^{-1}$
between 9~R$_*$ and 350~R$_*$.
To fit the observed line fluxes of $^{12}$CO, 
$^{13}$CO, H$_2$O and SiO with {\sc smmol} 
non-LTE line radiative transfer code, 
along with a mass-loss rate of $1.85\times10^{-4}$\,$M_{\odot}$\,yr$^{-1}$. 

To fit the high rotational lines of CO and H$_2$O, 
the model required a rather flat temperature distribution inside the dust 
condensation radius, attributed to the high H$_2$O opacity. Beyond the 
dust condensation radius the gas temperature is fitted best by an 
$r^{-0.5}$ radial dependence, consistent with the coolant lines becoming 
optically thin. Our H$_2$O emission line fits are consistent with an 
ortho:para ratio of 3 in the outflow.

\end{abstract}


\section{Introduction}

VY\,Canis\,Majoris is a high luminosity M-supergiant
\citep[2-3$\times10^5$\,$L_{\odot}$ for the parallax distance of
1.14$\pm$0.09\,kpc;][]{Choi:2008p27073} with a very high mass loss rate
\citep[$\sim2\times10^{-4}$\,$M_{\odot}$\,yr$^{-1}$;][]{Danchi:1994aj,DeBeck:2010fg}.
It is self-obscured by its dusty circumstellar envelope, which has
produced a reflection nebula at optical wavelengths
\citep{Humphreys:2007p26971}. Due to its brightness and high mass loss
rate, VY\,CMa has been studied extensively from optical to millimetre
wavelengths \citep[e.g.][]{Monnier:1999p27021}. Its far-infrared spectrum
shows numerous water lines \citep{Neufeld:1999p27088,
Polehampton:2010p28352}, while several water maser lines have been
discovered at submm wavelengths \citep[e.g.][]{Menten:2006p27227}. In
addition to CO, SiO, H$_2$O and other oxygen-rich molecules, carbon-rich
species such as HCN and CS have also been found in the spectrum of this
oxygen-rich object \citep{Ziurys:2007p27081}. 


The far-infrared and submillimetre spectra of red supergiants and
asymptotic giant branch (AGB) stars exhibit numerous molecular emission
lines \citep[e.g.][]{Barlow:1996tj, Royer:2010p29058}.
These can be used to investigate the physical and chemical state of
the circumstellar material, with different transitions probing specific
density and temperature domains. Water vapour rotational emission lines
are particularly strong in the far-infrared and submillimetre spectra of
cool oxygen-rich stars and can be important coolants of their stellar wind
outflows \citep[e.g.][]{Deguchi:1990p27374, Kaufman:1996p26003}. Following
the publication of our initial {\em Herschel} spectroscopic results for
VY\,CMa \citep{Royer:2010p29058}, we present here a more detailed analysis
of our SPIRE and PACS spectra of several molecular species, in particular
non-Local Thermodynamic Equilibrium (LTE) radiative line transfer models for observed CO, SiO and H$_2$O
transitions. This modelling has enabled us to find best-fitting radial
temperature and density profiles for VY~CMa's outflow, as well as allowing 
an estimate to be made of its $^{12}$C/$^{13}$C isotopic ratio.

\section{Observations and line identifications}

The {\em Herschel Space Observatory} (hereafter {\it Herschel}) was
launched in May 2009 \citep{Pilbratt:2010p29312}, with three instruments
on-board: SPIRE \citep{Griffin:2010p29303}, PACS
\citep{Poglitsch:2010p28964} and HIFI \citep{deGraauw:2010p29902}. We
report here observations made with the SPIRE Fourier Transform
Spectrometer (FTS) and with the PACS grating spectrometer, which together
cover the wavelength range from 55 to 650\,$\mu$m. The SPIRE FTS covers
the 190--650\,$\mu$m wavelength range, simultaneously, while the PACS
spectrometer covers the 55--210\,$\mu$m spectral range, similar to that
covered by the Long Wavelength Spectrometer (LWS) on-board the {\em
Infrared Space Observatory} \citep{Clegg:1996p26437}, although with higher
spectral resolution and greater sensitivity.

\begin{figure*}
\centering
\includegraphics[89,371][551,697]{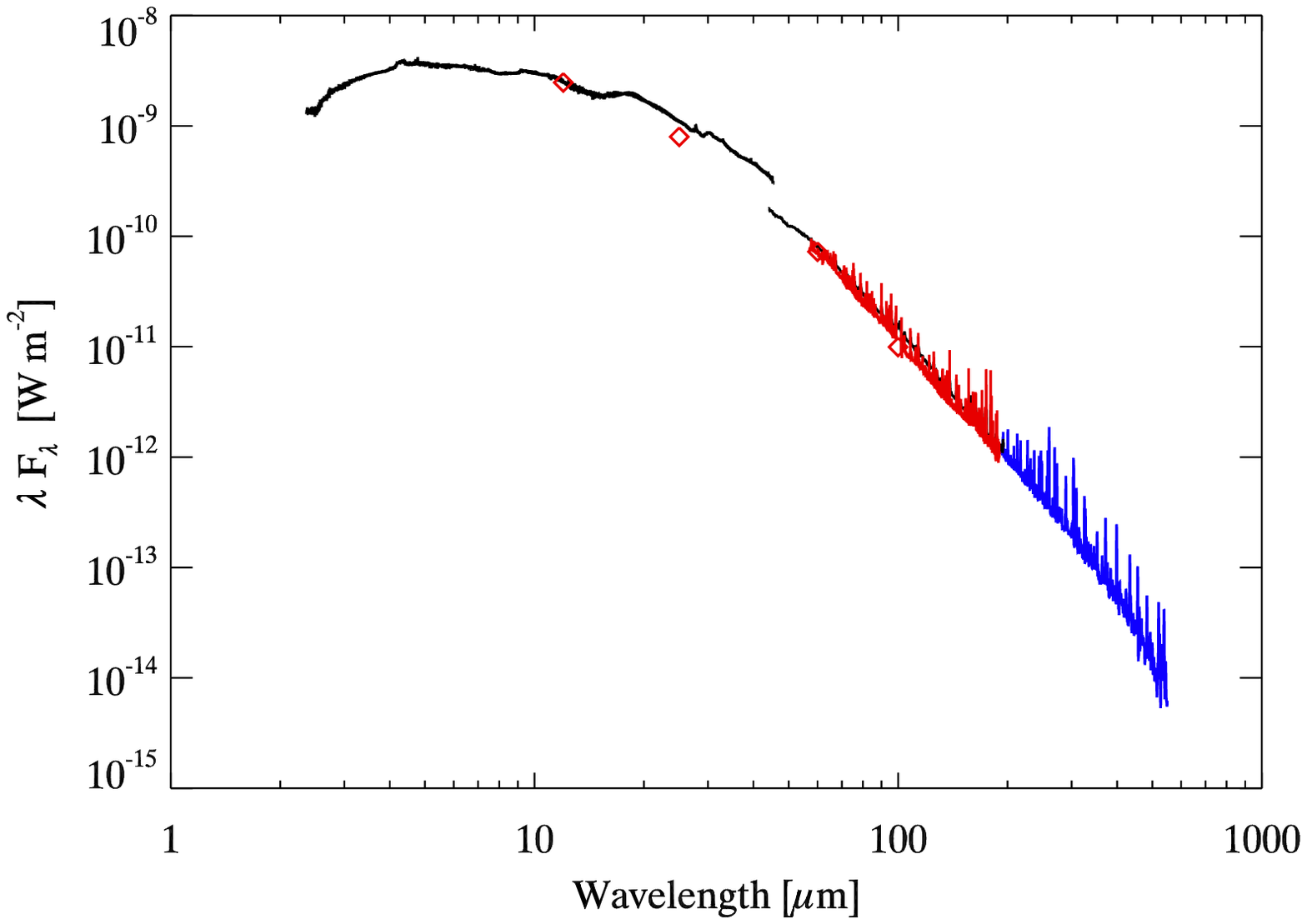}
\caption{The infrared spectral energy distribution of VY~CMa. The 
SPIRE FTS spectrum covers the 194--670\,$\mu$m range (blue), while 
the PACS spectrum covers the 55--190\,$\mu$m range (red). Additional 
data-points include the {\em ISO} SWS spectrum \citep{Sloan:2003p28681}, 
and the {\em ISO} LWS spectrum \citep{Polehampton:2010p28352}, both shown 
in black, and the {\em IRAS} point source catalog fluxes 
\citep{Beichman:1988p29907}, shown as red diamonds. 
Numerous emission lines are visible in the PACS and SPIRE spectra. 
}
\label{fig-sed}
\end{figure*}

Figure\,\ref{fig-sed} shows the observed spectral energy distribution of
VY\,CMa, including spectrophotometry from the ISO SWS and LWS instruments
(black) and the {\em Herschel} PACS (red) and SPIRE (blue) spectrometers.

\subsection{SPIRE FTS spectra}

The SPIRE FTS simultaneously covers the SLW short wavelength band 
(190--313\,$\mu$m; 31--52\,cm$^{-1}$; 957--1577\,GHz) and SLW long 
wavelength band (303--650\,$\mu$m; 15--33\,cm$^{-1}$; 461--989\,GHz). 
VY\,CMa was observed with the SPIRE FTS as part of the MESS (Mass-loss of 
Evolved StarS) Guaranteed Time Key Programme \citep{Groenewegen:2011hi}, 
first during HerschelÕs performance verification (PV) phase and again 
during nominal operations. The first SPIRE FTS spectrum was obtained on 
2009 September 13th (OD 123, obsid 1342183813). The on-source integration 
time was 3996\,s, corresponding to 60\,FTS scans. The second FTS spectrum 
was obtained on 2010 March 27th (OD 317, obsid 1342192834) with an 
on-source integration time of 2264\,s, corresponding to 34\,FTS scans.
  
We reduced the spectra with the first reliable version of the Herschel 
Interactive Processing Environment software \citep[version 4 of {\sc 
hipe;}][]{Fulton:2008bt} using the standard point-source pipeline 
described by \citet{Fulton:2010gt} and the calibration scheme described by 
\citet{Swinyard:2010p29297}. Since the initial data reduction an improved 
pipeline and calibration scheme has been evolved ({\sc hipe} v11, Fulton 
et al. in prep) which provides much superior calibration and noise 
reduction for faint sources.  VY ~CMa is a bright source for the SPIRE FTS 
and we find that comparison of the latest reduction with the reduction 
used in this paper shows no discernible difference in flux or noise 
levels.  The SPIRE FTS calibration is now based on the spectrum of Uranus 
and detailed analysis of the flux uncertainties (Swinyard et al. in prep) 
shows that the absolute flux on a source of this brightness is absolutely 
calibrated to within 6\%.
 
Fourier transform spectrometers produce spectra in which lines have sinc 
profile functions. To mitigate the effects on weak lines of negative 
side-lobes from many bright lines, we apodised the interferograms using 
the extended Norton-Beer function 1.5 \citep{Naylor:2007kl}. This yielded 
spectra in which the line profiles, which were close to Gaussian, had a 
full width half maximum (FWHM) of 0.072 cm$^{-1}$, compared to the 
unapodized spectral resolution of 0.048 cm$^{-1}$. Since fourier transform 
spectrometers produce spectra that are linearly sampled in frequency, all 
our FTS spectral measurements have been made in frequency space.

\subsection{PACS CO line spectra}

VY\,CMa was observed with the PACS spectrometer on November 3rd 2009 (OD 
173), using the full-range spectral energy distribution (SED) mode, with a
single repetition. The observations were taken in `chop-nodded'  mode, each
with a single nodding cycle, and using the smallest chopper throw available
(1 arcmin). The total duration of the observations was 3655\,sec for  the
54--73-$\mu$m and 102--210-$\mu$m ranges and 2759\,sec for the 67-109.5-$\mu$m
and 134--219-$\mu$m ranges. The data from the central spaxel  were extracted and 
reduced with
version 10 of {\sc hipe}  \citep{Ott:2010tla}, and calibration set 45. With
this set, the absolute flux calibration is based on comparisons of PACS'
internal calibration sources with celestial standards on the central 9
spaxels of the instrument. The calibration is established at one key
wavelength per camera and per band, and transferred to the other
wavelengths via a relative spectral response function calibrated on the
ground. After extraction and before `beam correction', a simple pointing
correction was applied to the flux in the central spaxel, by scaling it so
that its continuum matched the one measured in the integral of the central
3$\times$3 spaxels. This correction is part of the absolute flux calibration
scheme as it brings our spectrum to the same reference as used to
establish the absolute flux calibration files (i.e. the 9 central spaxels).
In our case, it is of the order of 15\% in all bands, and corrects for flux
missing from the central spaxel due to e.g. a slight mispointing, or pointing
jitter and/or marginal source extension. Additional details of the PACS
observations and data reduction are provided by \citet{Royer:2010p29058}.
The absolute flux calibration accuracy is estimated to be 20\,\%
\citep{Poglitsch:2010p28964}. For eight $^{12}$CO lines in the spectrum that 
had sufficiently good signal to noise ratios, Gaussian fits yielded the 
integrated line fluxes that are listed in Table~A1 of Appendix~A
(the PACS spectrum contains more than 400 molecular emission lines,
an overview of which is presented by \citet{Royer:2010p29058}).

\subsection{SPIRE line fluxes and identifications}

Prior to line identification and analysis, we measured the emission line
fluxes in continuum-subtracted versions of the SPIRE FTS spectra. The
continuum fits were obtained by fitting spline curves to connect
user-identified points that were judged to correspond to local continuum
levels through the spectra. Figure\,\ref{fig-spec1} shows the
continuum-subtracted SPIRE FTS spectra corresponding to the two different
observation dates (OD 123 and 317). The two spectra are almost
identical, although there are slight differences in the shapes of weak
features at the lowest frequencies ($<$20\,cm$^{-1}$ = 600 GHz). This is
because the continuum fluxes at these frequencies fall to levels where 
the flux uncertainties are larger than over the rest of the
spectrum. The spectral range between 939~GHz and 999~GHz
was covered by both the SSW and SLW subspectra.

The emission lines in the apodised SPIRE FTS spectra were fitted by 
Gaussian profiles using the Emission Line Fitting ({\sc elf}) suite of 
programs written by P. J. Storey that are part of the {\sc dipso} spectral 
analysis package mainly written by I. D. Howarth. The continuum-subtracted 
spectra were divided into smaller subspectra, for which about 20 emission 
lines at a time were fitted simultaneously with {\sc elf}. Since the lines 
were unresolved, all the lines in each sub-spectrum were required to have 
the same full width at half-maximum (FWHM). Over 260 emission lines were 
measured in total.

Amongst oxygen-rich AGB stars and red-supergiants, VY CMa is one of the
best studied at millimetre wavelengths \citep[e.g.][]{Tenenbaum:2010gu}.
Detected molecules include CS, HCN, HNC, CN, H$_2$S, NH$_3$, SiS, SO,
SO$_2$, TiO, TiO$_2$ \citep{Ziurys:2007p27081, Ziurys:2009p27100, Kaminski:2013gk}, in addition to
molecules commonly found in the spectra of oxygen-rich AGB stars, such as
CO, H$_2$O, OH, and SiO. 

Since most of the emission lines in the SPIRE FTS spectra are expected to 
be molecular transitions, we mainly searched molecular databases 
\citep[e.g.][]{Pickett:1991p29904, Muller:2005p29905}. Since there were 
multiple candidate transitions per observed line, in order to eliminate 
unlikely identifications we calculated molecular spectra based on an LTE 
(local thermodynamic equilibrium) radiative emission code 
\citep{Matsuura:2002p25132}, for species where Einstein coefficients or 
line intensities were available \citep[]{Rothman:2009p28626, 
Langhoff:1993p26503, Sauval:1984p26502, Muller:2005p29905}, namely CO, CN, 
CS, H$_2$O, H$_2$S, HCN, HCO$^+$, HNC, H$^3$O$^+$, OH, NH$_3$, SiO, SiS, 
SO, SO$_2$ and their isotopes, and compared the predicted line intensity 
patterns with the measured spectrum.

Tables\,\ref{linelist-1} and \ref{linelist-2} list for the detected lines 
in the SPIRE FTS spectra the measured line centre wavenumbers, integrated 
fluxes and the statistical uncertainties in these quantities as estimated 
by {\sc elf}, along with the rest frequencies (in cm$^{-1}$ and GHz) and 
wavelengths (in $\mu$m) of the proposed line identifications. These 
identifications are also indicated in Fig.\ref{fig-spec1}. Multiple 
identifications corresponding to more than three transitions are not 
plotted, but are indicated by vertical bars. The sources for the rest 
frequencies were the HITRAN database \citep{Rothman:2009p28626} for CO, 
$^{13}$CO, H$_2$O, H$_2^{18}$O, OH, SO$_2$, HCN, CS, NH$_3$ and H$_2$S, 
the JPL line database \citep{Pickett:1998jh} for HNC, SO, HCO$^+$, CN and 
H$_3$O$^+$ and the CDMS database \citep{Muller:2005p29905} for SiO. Our 
list of SPIRE FTS line identifications for the oxygen-rich outflow from 
VY~CMa complements those for three carbon-rich evolved sources that were 
presented by \citet{Wesson:2010p29903}.


\begin{figure*}
\centering
\resizebox{1.0\hsize}{!}{\includegraphics*[27, 68][565, 687]{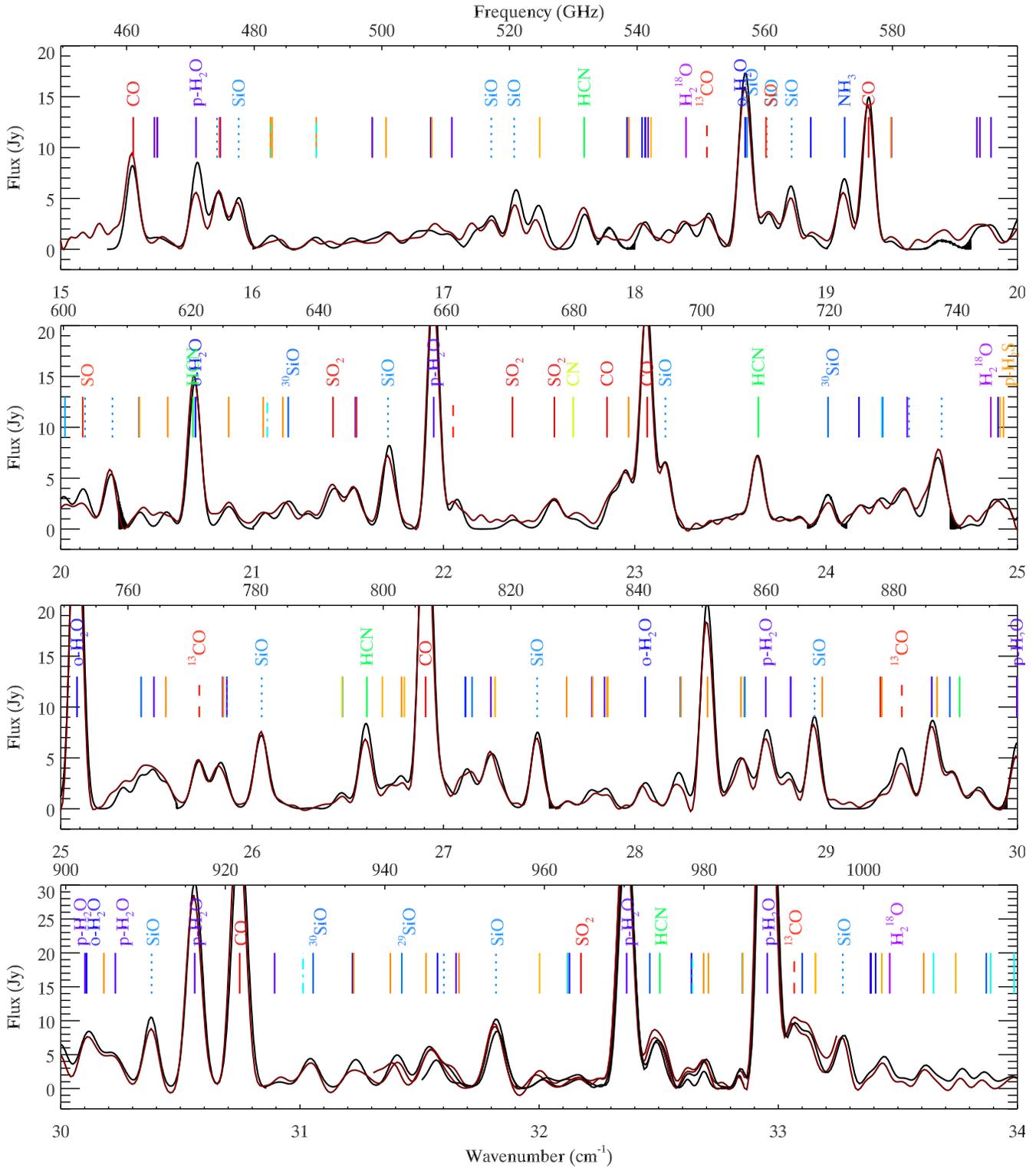}}
\caption{
The SPIRE FTS spectrum of VY\,CMa after continuum-subtraction. The black 
line shows the OD\,123 spectrum and the brown line shows the OD\,317 
spectrum.  Representative molecular identifications are labelled
with  different colour and line styles; occasionally vertical lines 
overlap. A full list of molecular line identifications can be found in 
Tables\,\ref{linelist-1} and \ref{linelist-2}.
The region between 31--33\,cm$^{-1}$ was covered by both the SLW and
SSW subspectra.
\label{fig-spec1}}
\end{figure*}
\addtocounter{figure}{-1}
\begin{figure*}
\resizebox{1.0\hsize}{!}{\includegraphics*[27, 68][565, 687]{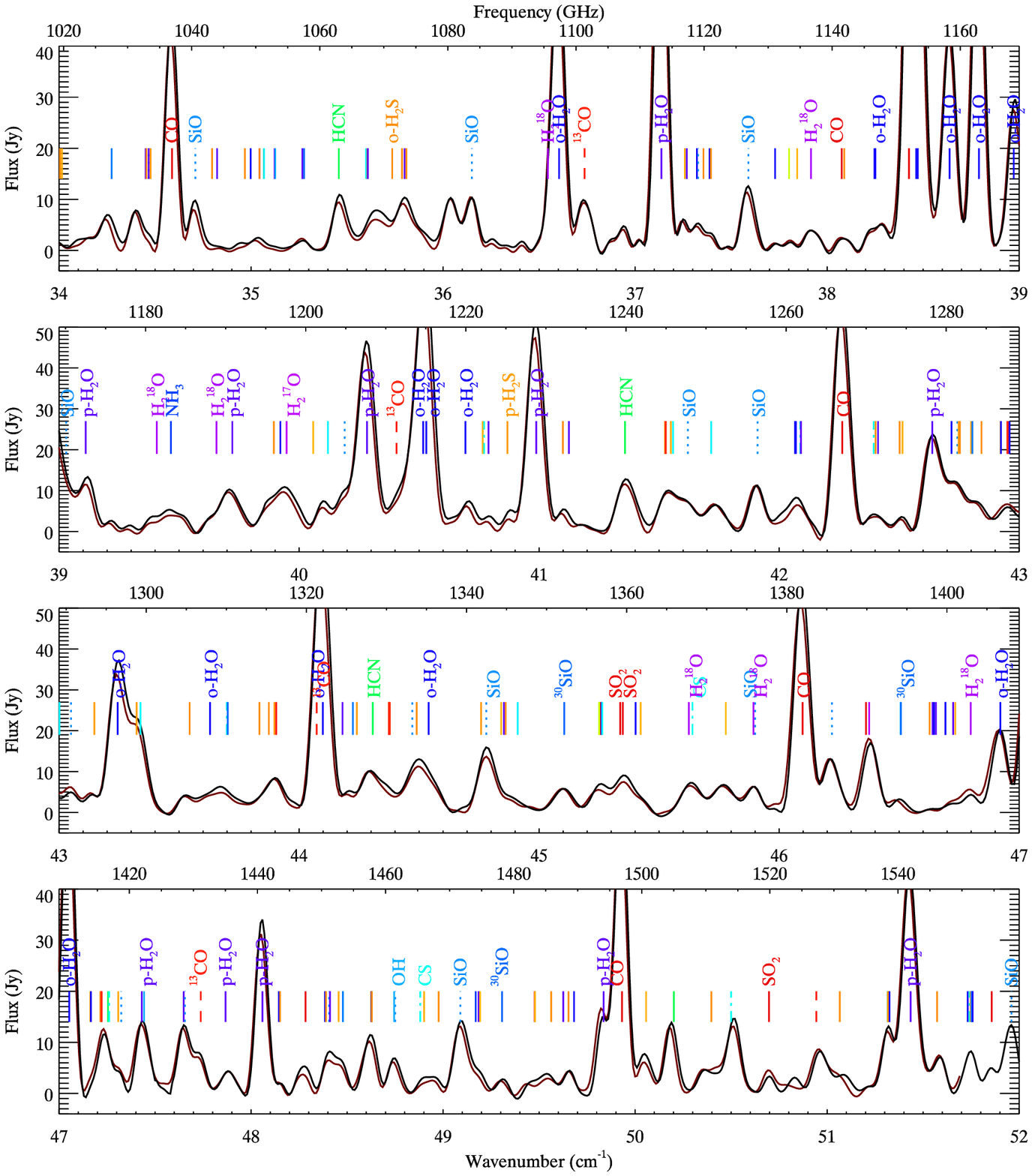}}
\caption{Continued.
\label{fig-spec2}}
\end{figure*}

Of the more than 260 lines in VY CMa's 460--1580~GHz spectral
region, listed in Tables\,\ref{linelist-1} and \ref{linelist-2}, nearly
one-third are attributed to H$_2$O, with the strongest lines in the
spectrum being pure rotational ($\nu_1, \nu_2, \nu_3$=0) transitions of water. Vibrationally excited
H$_2$O rotational lines, in particular $\nu_2$=1 transitions, are also
clearly detected. The pure rotational ($v$=0) lines of $^{12}$CO, $^{13}$CO and
HCN are strong and SiO rotational transitions also tend to be strong at
the longer wavelengths. SO and SO$_2$ contribute weaker lines; these
molecules have numerous transitions in this wavelength range, and could
potentially contribute many unresolved lines. Due to line blending, many
weak lines still do not have firm identifications.

\citet{Justtanont:2011wi} reported the detection of five SiO lines in the
{\em Herschel} HIFI spectra of nine oxygen-rich AGB stars. These included
pure rotational lines of SiO at 607.599 GHz (20.26\,cm$^{-1}$; $J$=14--13)
and 694.275 GHz (23.16\,cm$^{-1}$; $J$=16--15), along with vibrationally
excited ($v$=1) SiO rotational lines at 560.326\,GHz (18.69\,cm$^{-1}$;
$J$=13--12), 646.429\,GHz (21.56\,cm$^{-1}$; $J=$15--14), and 990.355\,GHz
(33.03\,cm$^{-1}$; $J$=23--22). These transitions and more are detected in
the VY\,CMa SPIRE spectra. Ground-based APEX observations have detected
$^{29}$SiO and $^{30}$SiO pure rotational lines in the submm spectra of 
VY~CMa \citep{Menten:2006p27227}. Transitions from these species are also
detected in the SPIRE FTS spectra.

\citet{Menten:2010p29072} reported the detection of the $1_{0}$--$0_{0}$ 
transition of ortho-NH$_3$ at 572.4981 GHz (19.096 cm$^{-1}$) from VY~CMa. 
This transition was detected as a relatively strong line in the 
SPIRE spectra. Other NH$_3$ lines are not obvious.

A number of far-infrared and submm H$_2$O lines have been predicted to
mase \citep{Deguchi:1990p27374, Neufeld:1991ky, Yates:1997p28231}. 
For VY\, CMa, \citet{Harwit:2010p29900} reported the
H$_2$O $5_{32}$--$4_{41}$ transition at 620.70\,GHz (20.70\,cm$^{-1}$) to
be masing, while \citet{Menten:2008p29901} found the H$_2$O
$6_{42}$--$5_{51}$ transition at 470.889\,GHz (15.70\,cm$^{-1}$) and the
$5_{33}$--$4_{40}$ transition at 474.680\,GHz (15.83\,cm$^{-1}$) to be
masing.  \citet{Menten:1995gt} reported 
the $1_{10}$--$1_{01}$ $\nu_2=1$ transition at 658.01\,GHz (21.95\,cm$^{-1}$) 
to be masing.
These lines were all detected in the SPIRE FTS spectra. However
their line fluxes are not particularly strong compared to those of other 
H$_2$O lines, suggesting that line strength alone may be 
an insufficient criterion to distinguish a maser line from a thermally 
excited line - we return to this question in Section 4.4.

\section{LTE emission line analysis}

In the case of optically thin LTE emission, a spectral line energy
distribution (e.g. Fig.\,\ref{fig-energy-CO}) can provide an estimate of
the excitation temperature and the number of emitting molecules, with the
upper level energy $E_{\rm u}$ at the peak of the overall curve providing
an estimate of the excitation temperature ($T_{\rm ex}$) of the molecules
\citep{Goldsmith:1999ia}. We adopted molecular constants from the HITRAN
database \citep{Rothman:2009p28626}, except for SiO, where we adopted
A-coefficients from \citet{Langhoff:1993p26503} and wavenumbers from
\citet{Muller:2005p29905}. Theoretical line intensities were then
calculated using an LTE radiative transfer code 
\citep{Matsuura:2002p25132}. The resultant fits are summarised in 
Table~\ref{table-LTE}.

\citet{Justtanont:2000p27249} and
\citet{Polehampton:2010p28352} used ISO/LWS measurements to estimate
excitation temperatures for CO. We note that a single line trend is found
only for  some simple diatomic and linear triatomic molecules. Water, for example, has three
rotational dipole moments, so its Einstein A-coefficients are more
complicated and its modified rotational diagram does not follow {\it a
single curve} relation \citep{Goldsmith:1999ia}. For CO, SiO and HCN, we
present our results in the form of observed and model spectral line energy
distributions, whereas for H$_2$O we compare observed with non-LTE model 
line spectra, in Sect.\ref{non-LTE}.

\begin{table}
  \caption{Results from LTE fits to the spectral line energy 
distributions. \label{table-LTE}}
\begin{center}
 \begin{tabular}{llrrrrrrrccccccccc}
\hline
Molecule & $T_{\rm ex}$ (K) & No. of molecules  & Mass ($M_\odot$)\\ 
\hline
CO              & 350$\pm$90   & $(1.2\pm0.4)\times10^{52}$  &  $(2.9\pm0.7)\times10^{-4}$ \\
$^{13}$CO  & 350$\pm$90   & $(1.8\pm0.9)\times10^{51}$  &  $(4.8\pm2.5)\times10^{-5}$ \\
SiO (hot)      & $\sim$1000    & $4\times10^{48}$                  &  $2\times10^{-7}$ \\
SiO (cold)    & 200                 & $2\times10^{48}$             &  $7\times10^{-8}$ \\
HCN            & 250$\pm$50         & $(2.7\pm2.1)\times10^{48}$   &  $(6.6\pm0.5)\times10^{-8}$\\
\hline 
\end{tabular}
\end{center}
\end{table}


\subsection{The $^{12}$CO lines}

\begin{figure}
\centering
\resizebox{1.0\hsize}{!}{\includegraphics*[67, 33][580, 428]{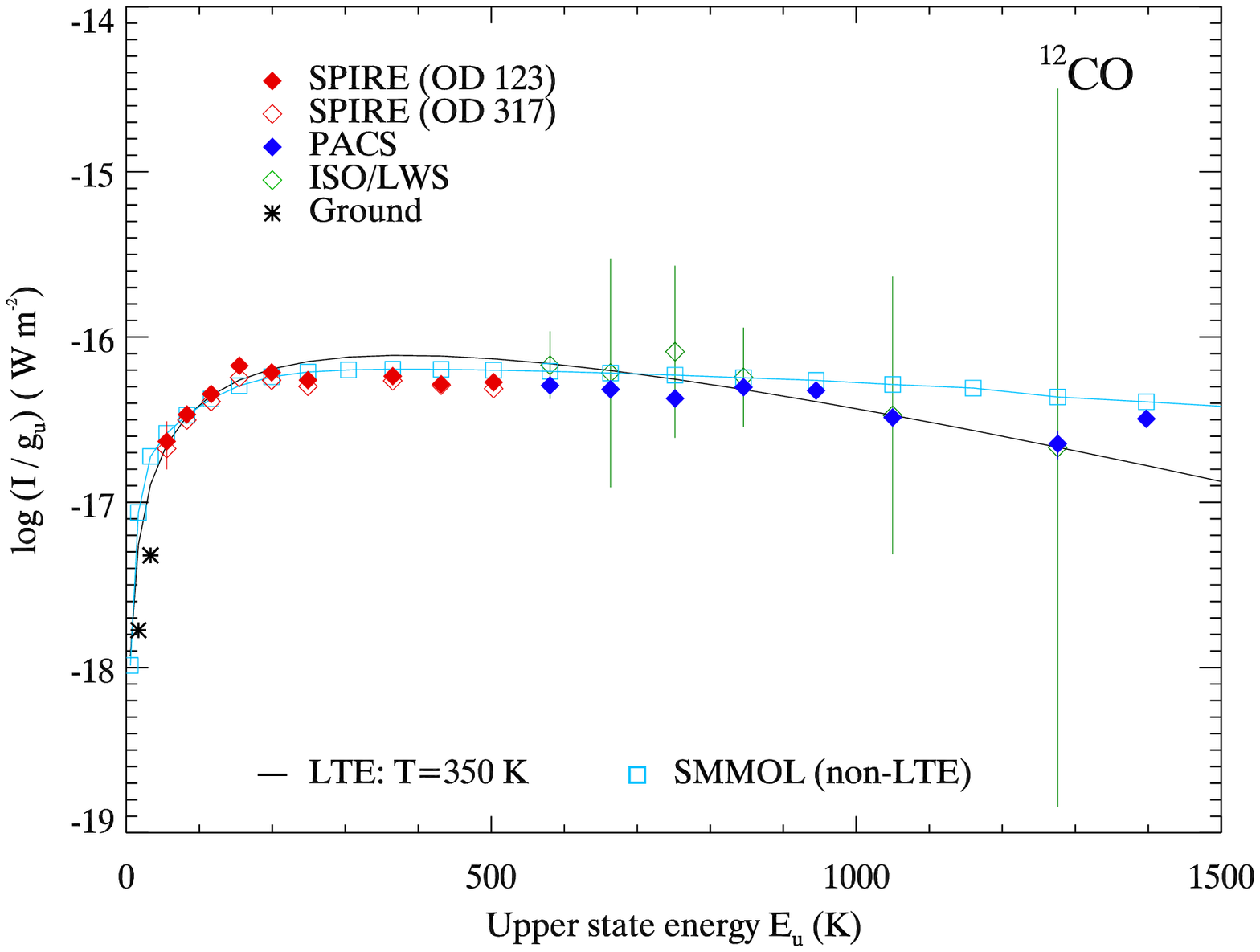}}
\caption{The spectral line energy distribution for VY~CMa's pure 
rotational $^{12}$CO 
transitions. As well as SPIRE and PACS spectrometer measurements, we have 
included ISO/LWS line flux data \citep{Polehampton:2010p28352} and the 
JCMT $J$=2--1 and 3--2 measurements of \citet{Kemper:2003kt}. The large 
error bars pertain to the ISO/LWS line flux measurements; those for the 
{\em Herschel} PACS and SPIRE line fluxes are smaller than the symbol 
sizes. We have also 
plotted model CO line intensities from an LTE model (solid line) and from 
the {\sc smmol} non-LTE model described in Section~4 (open squares). 
\label{fig-energy-CO}}
\end{figure}

Fig.\,\ref{fig-energy-CO} shows the $^{12}$CO spectral line energy
distribution for VY~CMa. The plotted line fluxes are from the SPIRE and
PACS spectra, along with the ISO/LWS line flux measurements of
\cite{Polehampton:2010p28352}. We have plotted the SPIRE line measurements
from both epochs. We have also included ground-based CO $J=$2--1 and 3--2
line flux measurements made with the James Clerk Maxwell Telescope
(JCMT) \citep{Kemper:2003kt}. The ground-based observations of these two
lines by \citet{Ziurys:2009p27100} gave line fluxes that are consistent
with those of \citet{Kemper:2003kt}.

The measured SPIRE, PACS and ISO/LWS $^{12}$CO line fluxes approximately
follow an LTE distribution that can be fitted with an excitation
temperature of 350$\pm$90~K, see Fig.\,\ref{fig-energy-CO}.
\citet{Polehampton:2010p28352} estimated a $^{12}$CO excitation
temperature of 250$\pm$140\,K, consistent with our value within the
respective uncertainties. We also modelled the $^{12}$CO line intensities
using the non-LTE code {\sc smmol}; details of the models are given in
Section~4.

Ground-based measurements of several $^{12}$CO lines in the {\em Herschel}
 range are available for VY~CMa. \citet{Kemper:2003kt} reported a CO
$J=6-5$ line flux of $1.4\times10^{-16}$\, W\,m$^{-2}$. They also observed
the CO $J=7-6$ line, with a line flux of $2.2\times10^{-16}$\,
W\,m$^{-2}$. These fluxes are a factor of 4--5 times smaller than our
SPIRE FTS measurements in Table\,\ref{linelist-1}. Kemper et al. used the
JCMT, for which the half-power beamwidth for the 6--5 and 7--6 lines was 8
and 6~arcsec, respectively. \citet{Muller:2007p26972} reported the angular
diameter of VY~CMa to be almost 10~arcsec in the CO $J$=2--1 line, so it is
possible that the JCMT observations of the higher-$J$ CO lines resolved
out some of their emission. The SPIRE FTS beam has a FWHM of
$\sim$37~arcsec in the 303--650-$\mu$m (15--33\,cm$^{-1}$; 461--989\,GHz) SLW region and $\sim$18~arcsec in
the 190--313-$\mu$m (31--52\,cm$^{-1}$; 957--1577\,GHz) SSW region \citep{Swinyard:2010p29297}, significantly
larger than that of the JCMT.

\subsection{The $^{13}$CO lines and the $^{12}$C/$^{13}$C ratio}

The $^{12}$C/$^{13}$C isotope ratio can be estimated from observations of
the lines of $^{12}$CO and $^{13}$CO. We measured a mean
$^{12}$CO/$^{13}$CO line flux ratio of 7.0$\pm$1.8 in the OD123 and in the
OD317 SPIRE FTS spectra (eight matched pairs of transitions in each
spectrum, from 5--4 to 13--12, with the blended 12--11 transition omitted).
If the transitions are all optically thin, then this ratio should be close
to the $^{12}$C/$^{13}$C ratio. Our LTE fits to the $^{12}$CO and
$^{13}$CO spectral line energy distributions (Fig.\,\ref{fig-energy-CO}
and Fig.\,\ref{fig-energy-13CO}) yielded an excitation temperature of
350$\pm$90~K and a $^{12}$C/$^{13}$C ratio of 6.2$\pm$2.0. Finally, our
best-fitting non-LTE model for $^{12}$CO and $^{13}$CO (Table
\ref{table-smmol-para}) implies a $^{12}$C/$^{13}$C ratio of 5.6$\pm$1.8 
for VY~CMa. 

Our $^{12}$C/$^{13}$C isotopic ratio of 5.6$\pm$1.8 is significantly lower 
than previously reported values for VY~CMa, but is within the range of the 
$^{12}$C/$^{13}$C ratios of 3--14 measured for four other red supergiants 
by \citet{Milam:2009p15019}, who derived a $^{12}$CO/$^{13}$CO ratio of 
25--46 for VY~CMa using the $J=$1--0 and 2--1 $^{12}$CO and $^{13}$CO 
lines. \citet{Nercessian:1989p27097} estimated a $^{12}$C/$^{13}$C ratio 
of 36$\pm$9, using the $J$=1--0 HCN and H$^{13}$CN lines. Our non-LTE 
models for $^{12}$CO and $^{13}$CO imply that their lowest transitions are 
optically thick, while the $^{12}$CO $J$=5--4 and higher transitions are optically thin. 
It is noticeable in the plots of \citet{DeBeck:2010fg} that while the 
profiles of the CO 2--1 and 3--2 transitions show several pronounced peaks 
and troughs, the 4--3, 6--5 and 7--6 transitions show increasingly 
Gaussian-like line profiles; \citet{DeBeck:2010fg} derived a 
$^{12}$C/$^{13}$C of 14.8, much closer to the ratio derived from the 4--3 
through to 13--12 lines in the SPIRE-FTS spectra than those obtained by 
observers making use of lower-J lines. We 
believe that our $^{12}$C/$^{13}$C ratio of 5.6$\pm$1.8 for VY~CMa is 
reliable and removes an anomaly by yielding an isotopic ratio that is 
consistent with those found for other red supergiants.

\subsection{HCN}

Emission lines from the carbon-bearing molecule HCN are predominantly
found in the spectra of carbon-rich AGB stars but have also been found in
the spectra of several oxygen-rich AGB stars at millimeter wavelengths
\citep[e.g.][]{Deguchi:1985kb, Nercessian:1989p27097}. HCN emission was
detected from the F8 supergiant IRC+10420 by \citet{Jewell:86}, while
\citet{Ziurys:2009p27100} detected HCN, HNC, CN and CS emission from
VY~CMa. Over ten HCN lines have been detected in our {\em Herschel} FTS
spectra of VY~CMa, where this molecule is responsible for lines of
intermediate strength. Figure\,\ref{fig-energy-HCN} shows the spectral
line energy distribution for HCN obtained from the SPIRE FTS data. It has
been fitted with an LTE model having $T_{\rm ex}$ = 250$\pm50$\,K, lower than
that derived for CO, which could indicate that HCN originates from cooler
outer parts of the circumstellar envelope.

\citet{Nercessian:1989p27097}, \citet{Charnley:1995wo} and
\citet{Willacy:1997p27138} have investigated processes which can trigger
the formation of carbon-bearing molecules in oxygen-rich circumstellar
envelope. \citet{Charnley:1995wo} and \citet{Willacy:1997p27138} used
CH$_4$ as a parent molecule and formed carbon-bearing molecules through
photochemical processes mediated by the interstellar UV radiation field. 
Their models predicted a peak fractional
abundance of HCN at about $10^{16}$--$10^{17}$\,cm in radius. This is in
contrast with CO, which was predicted to have an almost constant abundance
from $10^{15}$\,cm out to $10^{18}$\,cm. Their models were for AGB stars,
so for red supergiants, the overall sizes would have to be scaled up. The
predicted lack of HCN in the inner part of the outflow could be
responsible for the lower HCN excitation temperature compared
to that of CO. In contrast,
\citet{Duari:1999p27058} investigated shock chemistry in the extended
atmosphere, i.e. just above the photosphere. This can also form carbon-bearing
molecules in oxygen-rich environments and they favoured shock chemistry to
account for the HCN observed around IK\,Tau. Shock formation of HCN in the
innermost parts of an outflow, where CO also exists, would be expected to lead 
to similar HCN and CO spectral line energy distributions, or to even give HCN 
a higher excitation temperature than CO. This is not in agreement with our SPIRE
FTS observations of CO (Fig.\,\ref{fig-energy-CO}) and HCN 
(Fig.\,\ref{fig-energy-HCN}), so for HCN this 
appears to favour an origin in the cooler outer parts of the outflow,
i.e. via photochemical processes mediated by the interstellar
radiation field.

\subsection{SiO}

More than twenty SiO lines were detected in the SPIRE spectra. The
spectral line energy distribution for the pure-rotational transitions  in $v=0$  of
SiO shows a double peak (Fig.\,\ref{fig-energy-SiO}): the first peak is at
about $E_{\rm u} \sim$ 100--200\,K, with the line intensities dipping
until about $E_{\rm u}\sim500$\,K and then rising again for $E_{\rm
u}>$500\,K. The fits to these lines suggest a cool component with $T_{\rm
ex}$=200\,K and a hotter component with $T_{\rm ex} \sim$ 1000\,K. The
latter value is rather uncertain, because its peak $E_{\rm u}$ is not well
constrained.

These two SiO components could reflect the chemistry and structure of
VY\,CMa's outflow. Signatures of deep regions of its atmosphere have been
found in its near- and mid-infrared molecular emission
\citep[e.g][]{TSUJI:1997ws}. The excitation temperature of molecular gas
associated with the extended atmosphere and inner envelope is typically
500--1500\,K \citep[e.g][]{TSUJI:1997ws, Ireland:2011ij}, similar to the
excitation temperature found here for the hotter SiO component
($\sim$1000\,K), so it appears likely that the hot SiO component is
associated with gas in the inner outflow. Further out in the extended
envelope dust forms, and radiation pressure on dust grains can drive the
outflow, creating the circumstellar envelope from which the cooler SiO
component originates.

\begin{figure}
\centering
\resizebox{1.0\hsize}{!}{\includegraphics*[67, 33][580, 428]{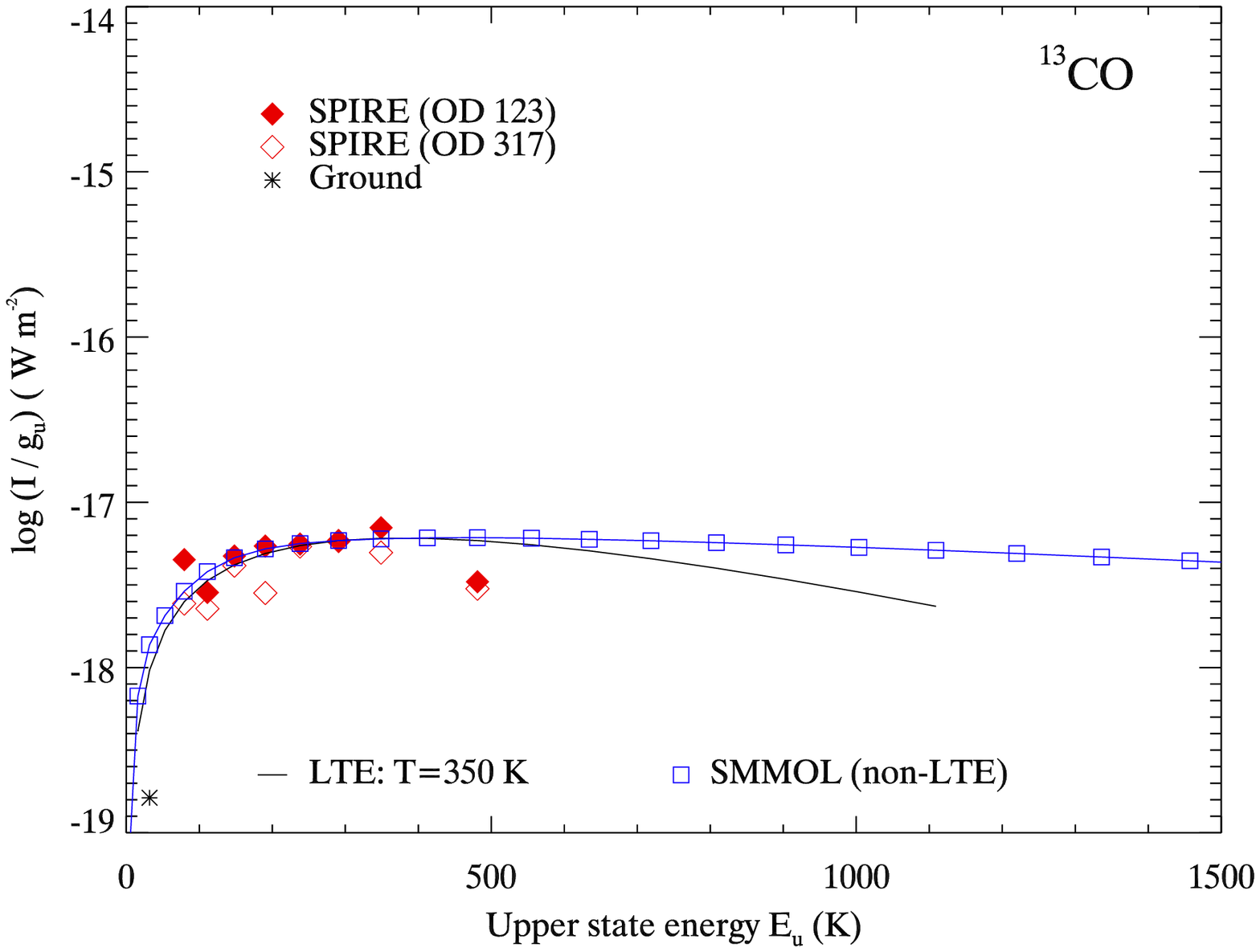}}
\caption{The spectral line energy distribution for the pure rotational transitions of $^{13}$CO 
measured in the SPIRE FTS spectra of VY~CMa.  
\label{fig-energy-13CO}}
\end{figure}
\begin{figure}
\centering
\resizebox{1.0\hsize}{!}{\includegraphics*[67, 33][580, 428]{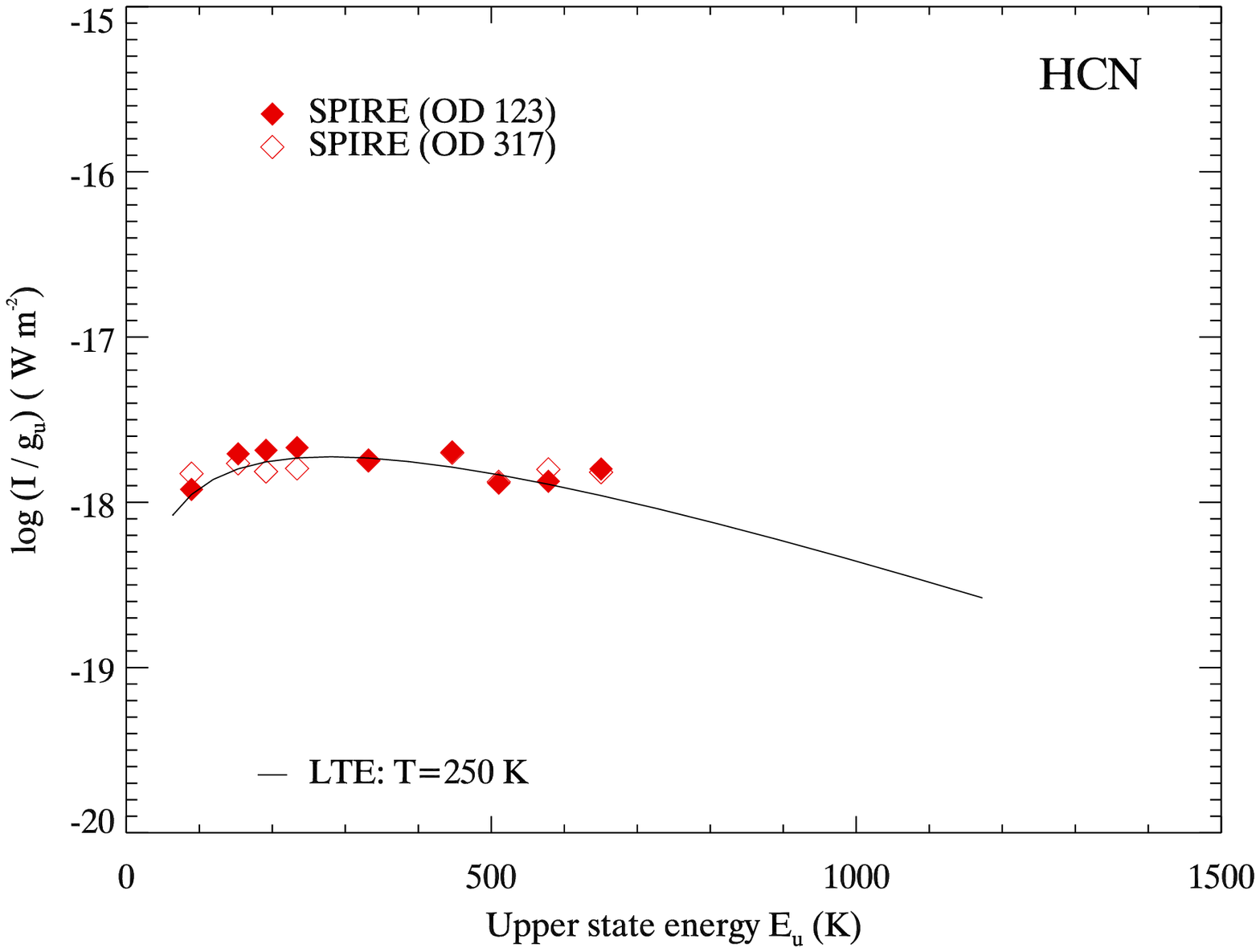}}
\caption{The spectral line energy distribution for the pure rotational transitions of HCN measured 
in the SPIRE FTS spectra of VY~CMa. 
\label{fig-energy-HCN}}
\end{figure}

\begin{figure}
\centering
\resizebox{1.0\hsize}{!}{\includegraphics*[67, 33][580, 428]{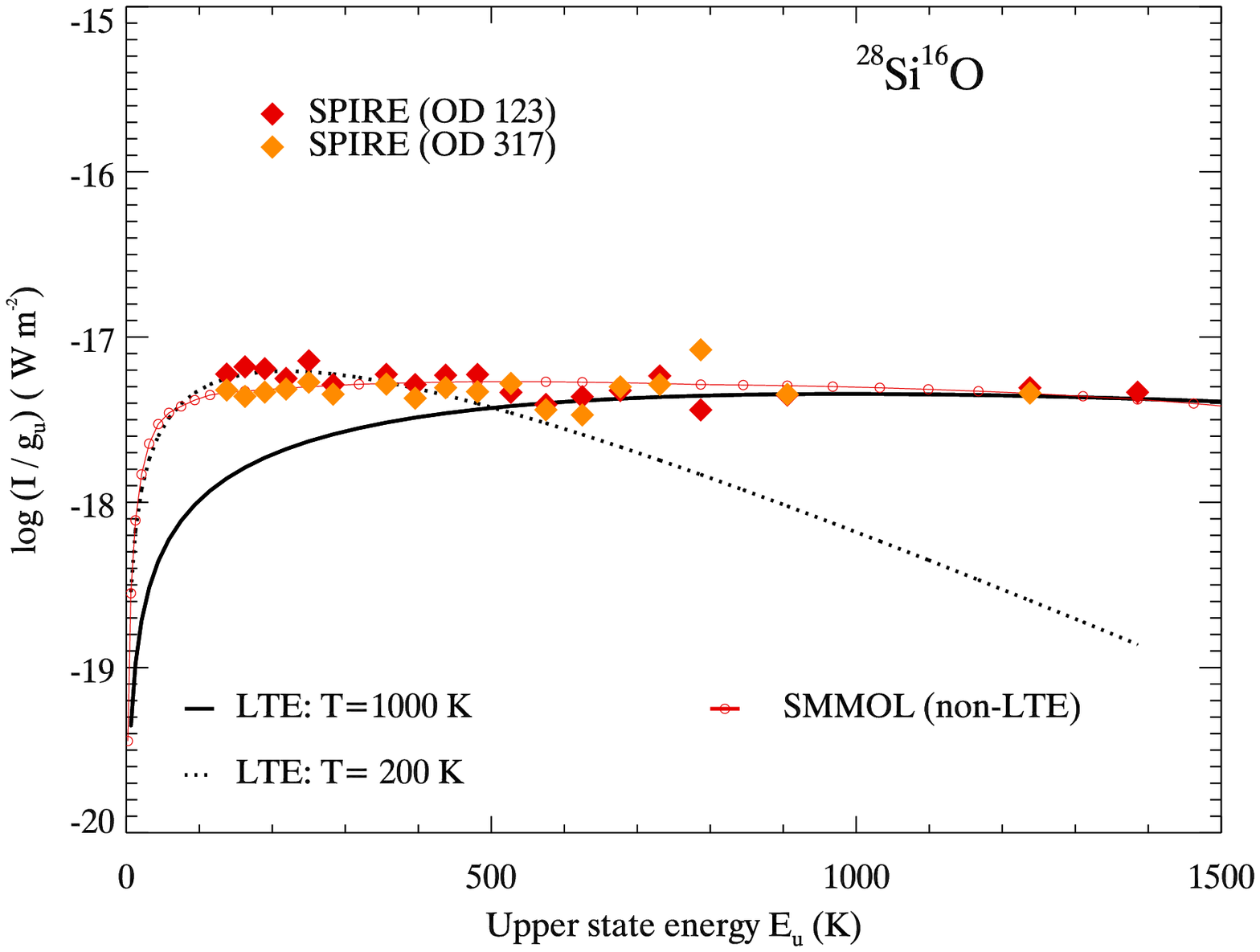}}
\caption{The spectral line energy distribution of the pure rotational transitions of SiO 
measured in the SPIRE FTS spectra of VY~CMa. 
\label{fig-energy-SiO}}
\end{figure}

\section{Non-LTE model analysis}\label{non-LTE}

The far-IR and submm spectra of VYCMa display a host of molecular
emission lines, many of which are due to water vapour.
Some of these molecular 
lines may have significant optical depths and are most likely not in LTE, 
so that non-LTE radiative transfer codes are required to interpret the
physical conditions in the line emitting regions.
We have modelled the {\it Herschel} spectra of VY CMa using the non-LTE 
code {\sc smmol} \citep{Rawlings:2001p28230}.

\subsection{Basic parameters}

VY\,CMa has been studied extensively and its basic parameters are 
well-constrained. The parallax measurements by \citet{Choi:2008p27073} 
obtained a distance of 1.14$\pm$0.09\,kpc and we have used this distance 
here. The effective temperature is adopted to be 2800\,K 
\citep{LeSidaner:1996p8500} and we adopt a stellar radius of $R_* = 
2069~R_\odot$, corresponding to a luminosity of 
2.37$\times10^5$~$L_\odot$. Table \ref{table-smmol-para} lists the main 
parameters used for our {\sc smmol} wind models.

\begin{table*}
 \caption{Parameters of the non-LTE model that produced the best
fits to the {\em Herschel} $^{12}$CO, $^{13}$CO, SiO and H$_2$O line fluxes 
of VY~CMa.
 \label{table-smmol-para}}
\begin{center}
\begin{tabular}{lrllllrrrrrrccccccccc}
\hline
Model parameters &  &  $^{12}$CO &  $^{13}$CO & H$_2$O & SiO \\ \hline
Stellar radius R$_*$ (cm)                                          & 1.44$\times 10^{14}$     \\
$R_{inner}$: inner radius of molecular gas envelope (cm)           & 1.44$\times 10^{14}$     \\
$R_{inner, dust}$: inner radius of dust envelope (cm)              & 1.283$\times 10^{15}$    \\
$R_{outflow, break}$: radius of density discontinuity (cm)         & 5.0$\times 10^{16}$     \\
$R_{outer, dust}$: outer radius of model (cm)                      & 2.93$\times 10^{17}$ \\
$^\ddag$$\beta$: density law index                                         & 2.0                      \\
$^\ddag$$\alpha$: Kinetic temperature law index                            & 0.6                      \\
Turbulent velocity (km\,s$^{-1}$)                                  & 1.0                      \\
X: fraction of molecule/H$_2$                                      & & 2.5$\times 10^{-4}$   & 4.5$\times 10^{-5}$ & 2$\times 10^{-4}$ & 8$\times 10^{-5}$ \\
Dust optical depth in the $V$-band                                 & 50                      \\
$\rho_d$: density of dust (g cm$^{-3}$)                            & 3.0                     \\
Gas to dust mass ratio: $\rho_g / \rho_d$                          & 267                     \\
Wind velocities: $v_{\infty}$ and $v_{inner}$ (km\,s$^{-1}$)       & 44.0, 4.0              \\
Inner radius of velocity law (cm)                                  &  1.283$\times 10^{15}$     \\
$^\ddag$$\gamma$: velocity law index                               &  0.2                     \\
Stellar temperature (K)                                            & 2800                    \\
Sublimation temp. of molecule (K)                                  & & 20                    & 20 & 100 & (1000)/100 \\
Mass-loss rate between $R_{inner, dust}$ and $R_{outflow, break}$ ($M_{\odot}$\,yr$^{-1}$)     &  $1.85\times10^{-4}$     \\
Mass-loss rate beyond $R_{outflow, break}$ ($M_{\odot}$\,yr$^{-1}$)                            &  $9.3\times10^{-5}$     \\
Dust emissivity behaviour $\kappa$($>$250\,$\mu$m)    & $\lambda^{-1}$          \\
H$_2$O ortho:para ratio                                            &        &   &            & 3:1 \\
$R_{T, break}$: Radius of break in temperature (cm)                   & 5$\times 10^{14}$ \\
New kinetic temperature law index $\alpha$ at $R_{T, break}$          &0.15 \\
SiO density reduction factor at $R_{inner, dust}$$^\dag$  & &&&& 20\\
\hline
\end{tabular}
\end{center}
$^\ddag$Gas temperature, density and velocity laws used for the 
non-LTE line modelling: Temperature: $(r/R_{inner})^{-\alpha}$,
Gas density: $(r/R_{inner})^{-\beta}$, \newline
Velocity:  $v(r) = v_{inner} + (v_{\infty} - v_{inner})$  $(1- R_{inner, dust}/r)^{\gamma} $, for $r \geq R_{inner, dust}$.\\
$^\dag$  The SiO density reduced by a factor 20 at $R_{inner, dust}$, 
attributed to dust condensation, with residual SiO extending out to a 
radius corresponding to a gas temperature of 20~K.
\end{table*}

The first task was to construct a dust radiative transfer model to fit the 
observed SED, in order to estimate the dust grain properties, the inner 
dust condensation radius and the radial density and temperature 
distributions of the dust grains.
This was calculated with a one-dimensional and spherically symmetric
model using the {\sc dart} dust radiative transfer code of
\citet{Efstathiou:1990p27044},  which result was also compared with Monte Carlo radiative transfer code
\citep{Ercolano:2005p29908}.
The dust was assumed to consist of
astronomical silicates \citep{Draine:1984jw} with an MRN 
(Mathis, Rumpl and Nordsieck) size distribution
of the form $n(a) da \propto a^{-3.5} da$, for grains with radii $a$
between 0.005-0.25$\mu$m \citep{mathis77}. Although the MRN distribution
was derived for interstellar grains, similar power-law distributions are
expected wherever grain-grain collisions determine their size
distributions, including stellar outflows \citep{biermann80}. A best fit
was found to the SED of VY\,CMa and the resulting dust temperature radial
profile was used as an initial input into the line radiative transfer
calculations. The fit required a dust condensation inner radius of
9~$R_*$ and an outer radius of at least 830~$R_*$.

The mass-loss structure and history of VY\,CMa has been discussed by
\citet{Decin:2006p27245} and by \citet{Muller:2007p26972}.
\citeauthor{Decin:2006p27245} used a spherically symmetric wind model with
line radiative transfer to interpret the ground-based CO spectra and
deduced that it had undergone a phase of high mass-loss
($\sim3.2\times10^{-4} M_{\odot}$\,yr$^{-1}$) some 1000\,yr ago and
lasting for some 100~yrs, preceded by a lower mass loss rate phase lasting
some 800 yrs. \citeauthor{Muller:2007p26972} deduced that the inner
part of VY~CMa's envelope has been undergoing a phase of enhanced mass
loss that began about 100~yrs ago. 
Our adopted terminal velocity of 44~km~s$^{-1}$ for the outflow is based on half the full widths at zero
intensity of CO line profiles presented by \citet{Kemper:2003kt}, as well as on expansion velocities 
determined by \citet{Richards:1998p27089, Menten:2008p29901} from H$_2$O masers.

\subsection{NLTE Line Radiative Transfer Calculations}

The molecular line radiative transfer and level population code {\sc 
smmol} \citep{Rawlings:2001p28230} uses the accelerated Lambda 
iteration (ALI) scheme described by \citet{Scharmer:1985p28621} and 
\citet{Rybicki:1991p28622} to solve the coupled level population and line 
radiative transfer problem exactly and quickly. These methods, designed to 
solve line radiative transport in optically thick stellar atmospheres, are 
ideal for solving the radiative transport by water molecules in the 
outflows from AGB stars and red supergiants. In these outflows it is clear 
that many ground vibrational state transitions of water and CO are 
optically thick at their line centres and accelerated $\Lambda$-iteration 
({\it ALI}) methods are needed to achieve an exact converged solution for 
the level populations and radiation field.

{\sc smmol} uses a fine grid in position-velocity space to allow the
inclusion of the effects of maser emission in the calculation of level
populations. This is particularly important for an outflow such as that of
VY\,CMa, which has very bright H$_2$O maser emission lines at submm
wavelengths \citep{Menten:2008p29901, Harwit:2010p29900}, which can
saturate and affect level populations. In a paper on Orion-KL H$_2$O line
emission \citep{Lerate:2010cj}, we used approximate
H$_2$$^{16}$O--H$_{2}$ collisional parameters based on H$_2$O--He
cross-sections \citep{Green:1993p28623}. Here we use H$_2^{16}$O--H$_2$
cross-sections calculated by \citet{Faure:2008p28232}, along with water
vapour molecular data from the the H$_2$O line list (containing half a
billion lines) produced by \cite{Barber:2006p29886}. The data include the
45 lowest rotational energy levels in the ground vibrational states of
both ortho- and para-H$_2$O, colliding with both para- and ortho-H$_2$,
covering a kinetic temperature range of 20\,K to 2000\,K.

\begin{table*}
  \caption{Reduced $\chi^2$ values for the {\sc smmol} model fits to 
the observed lines, together with mean and median $I_{obs}/I_{SMMOL}$ 
ratios.
\label{table-chi-sq}}
\begin{center}
 \begin{tabular}{ll lll lll lll}
\hline
\multicolumn{2}{c}{Molecules}
& $\chi^2_{red}$ & $\chi^2_{red}$ & $\chi^2_{red}$
& \multicolumn{3}{c}{mean of ${\frac{I_{obs}}{I_{SMMOL}}}$}
& \multicolumn{3}{c}{median of ${\frac{I_{obs}}{I_{SMMOL}}}$} \\
             & & \multicolumn{2}{c}{SPIRE} & PACS \\
             & & OD 123 & OD 317 & OD 173 & OD 123 & OD 317 &  OD 173  & OD 123 & OD 317 &  OD 173 \\ \hline
\hline
CO & SPIRE only                       & 0.12 & 0.26 &      & 1.03$\pm$0.21 & 0.97$\pm$0.30 &   & 1.01 & 0.93 \\
   & PACS only                        &      &      & 0.94 &               &               & 0.65$\pm$0.11 & & & 0.69       \\
   & SPIRE + PACS (OD 173)            & 0.27 & 0.30 &      & 0.86$\pm$0.26 & 0.83$\pm$0.28 &   & 0.84 & 0.80 \\
\multicolumn{2}{l}{$^{13}$CO}         & 0.67 & 0.83 &      & 1.05$\pm$0.62 & 0.84$\pm$0.55 &   & 1.15 & 0.80 \\
\multicolumn{2}{l}{SiO}               & 0.16 & 0.34 &      & 1.01$\pm$0.31 & 0.86$\pm$0.35 &   & 1.10 & 0.94 \\
\multicolumn{2}{l}{Para H$_2$O}       & 4.34 & 4.57 &      & 1.00$\pm$0.56 & 0.92$\pm$0.51 &   & 0.93 & 0.86 \\
\multicolumn{2}{l}{Ortho H$_2$O}      & 4.96 & 3.19 &      & 1.54$\pm$1.20 & 1.60$\pm$1.12 &   & 1.40 & 1.47 \\
\hline
\end{tabular}
\end{center}
\end{table*}


For $^{12}$CO and $^{13}$CO we use the CO--H$_{2}$ cross-sections
calculated by \citet{Yang:2010bb}, with molecular data from
\citet{Muller:2005p29905}. This provided CO energy level and collisional
data up to $J$=40, for kinetic temperatures ranging from 2\,K to 3000\,K.
These collisional rates were for collisions with both para- and
ortho-H$_2$.

The SiO--H$_2$ collisional excitation rates were based on the SiO--He
calculations of \citet{Dayou:2006ey}, scaled by a factor of 1.38 in order
to account for the difference between the He and H$_2$ cross-sectional
areas. The calculations included  41 rotational transitions only,
ignoring vibrationally excited transitions, which could potentially cause 
errors
for molecular lines arising from high temperature gas, typically over 
1000\,K.
These errors could particularly influence line intensities from 
SiO molecules located inside the dust forming radius. The 
inclusion of vibrationally excited levels will be the subject
of future work.

\subsection{Fitting the molecular emission line fluxes}

Having obtained estimates for the inner radius and radial density and 
temperature variations of the dust (Section 4.1), these were used as the 
starting points for modelling the molecular gas using {\sc smmol}, 
iterating until an overall fit to the $^{12}$CO and $^{13}$CO line 
fluxes was obtained, primarily by varying the gas density $n$(H$_2$)($r$) 
and the gas radial temperature and velocity profiles, $T(r)$ and $v(r)$. 

The radius at which H$_2$O molecules condense onto dust grains, i.e. below 
the sublimation temperature of water-ice, was estimated to be 
6.9$\times10^{16}$~cm, well inside the outer radius of 
2.9$\times10^{17}$~cm adopted for our model, so the depletion of gaseous 
H$_2$O by this effect was incorporated into the modelling. CO has a much 
lower sublimation temperature, so depletion of CO onto grains did not have to 
be taken into account. The penetration of interstellar UV photons can 
dissociate molecules in the outer regions of the outflow. To estimate the 
water dissociation radius the method described by \citet{Decin:2006p27245} 
was used, while for CO the approach followed by \citet{Mamon:1988br} was 
used. For CO we found that due to shielding by dust in the outflow the 
dissociation radius was outside the outer radius of 2.9$\times10^{17}$~cm 
of the model. The predicted non-dissociation of CO is consistent with the 
non-detection in our SPIRE FTS spectrum of the atomic carbon [C~{\sc i}] 
lines at 492 and 809~GHz. The predicted dissociation radius for H$_2$O
molecules was 2.0$\times10^{17}$~cm, just inside the outer radius
of the model but this did not have to be taken into account since 
all the gaseous water molecules had already been depleted onto dust
grains beyond 6.9$\times10^{16}$~cm.

A turbulent velocity of 1.0 km\,s$^{-1}$ was adopted based upon MERLIN 
observations of water maser clouds in VY\,CMa, which suggest a turbulent 
width of 1.0 km\,s$^{-1}$ \citep{Richards:1998p27089}. The initial wind 
radial velocity law was based upon maser observations 
\citep{Richards:1998p27089} and CO line observations \citep[and references 
therein]{Decin:2006p27245}. We constrained the outflow model first using 
the CO observations and the best-fitting CO model was then used as the 
starting point for the modelling of the H$_2$O and SiO line data, with 
$T(r)$ and $v(r)$ again being varied until a best fit was obtained for all of 
the molecular species. The structure of the outflow found from the 
best-fitting CO and H$_2$O models did indeed match.

\begin{table*}
  \caption{ Observed and predicted H$_2$O line fluxes in the SPIRE wavelength range, with model predictions for population inversion
and masing in these lines. 
   \label{table-smmol-maser}}
\begin{center}
 \begin{tabular}{rrr cr rrrcccccccccccccc}
\hline
\multicolumn{3}{c}{Spectral line} &Transition & $E_{\rm up}$ &  \multicolumn{2}{c}{Obs Int.$^{\S}$} &   \multicolumn{2}{c}{{\sc SMMOL}} & \multicolumn{2}{c}{Obs/Pred.$^{\P}$} \\
$\nu_0$ & $\nu_0$ & $\lambda_0$ & $J^\prime$$_{K_a^\prime}$$_{K_c^{\prime}}$--$J_{K_a}$$_{K_c}$ && $I_{obs}$ (OD 123) & $I_{obs}$ (OD 317) &$I_{\sc SMMOL}^{\dag}$ & Inverted?$^{\ddag}$ & \multicolumn{2}{c}{$I_{obs}$/$I_{\sc SMMOL}$}\\
       GHz  &     cm$^{-1}$  & $\mu$m  &  & cm$^{-1}$ & $10^{-18}$\,W\,cm$^{-2}$ &  $10^{-18}$\,W\,cm$^{-2}$& && OD 123 & OD 317  \\\hline
\multicolumn{10}{l}{{\it Para-H$_2$O}} \\
   470.8958  &       15.707 &   636.66 & 6$_{42}$-5$_{51}$  &  757.78 &  221.8 &  135.7 &   31.1 & Y    & 7.1 & 4.4\\
   474.7033  &       15.834 &   631.55 & 5$_{33}$-4$_{40}$  &  503.97 &  159.5 &  137.9 &   90.8 & Y    & 1.8 & 1.5\\
   752.0483  &       25.086 &   398.64 & 2$_{11}$-2$_{02}$  &   95.18 &  959.7 &  769.8 & 3503.0 & N    & 0.3 & 0.2\\
   906.2355  &       30.229 &   330.82 & 9$_{28}$-8$_{35}$  & 1080.39 &  196.2 &  131.1 &  151.9 & Y    & 1.3 & 0.9\\
   916.1888  &       30.561 &   327.23 & 4$_{22}$-3$_{31}$  &  315.78 &  788.2 &  664.4 & 1045.5 & Y    & 0.8 & 0.6\\
   970.3327  &       32.367 &   308.97 & 5$_{24}$-4$_{31}$  &  416.21 &  871.6 &  999.3 &  727.8 & Y    & 1.2 & 1.4\\
   987.9609  &       32.955 &   303.45 & 2$_{02}$-1$_{11}$  &   70.09 & 1912.0 & 1856.0 & 2533.3 & N    & 0.8 & 0.7\\
  1113.3673  &       37.138 &   269.27 & 1$_{11}$-0$_{00}$  &   37.14 & 2014.0 & 2040.0 & 3262.7 & N    & 0.6 & 0.6\\
  1172.5479  &       39.112 &   255.68 & 7$_{44}$-6$_{51}$  &  927.74 &  304.4 &  291.0 &  327.7 & Y(w) & 0.9 & 0.9\\
  1190.8655  &       39.723 &   251.75 & 8$_{53}$-7$_{62}$  & 1255.91 &  260.3 &  229.6 &  136.7 & N    & 1.9 & 1.7\\
  1207.6845  &       40.284 &   248.24 & 4$_{22}$-4$_{13}$  &  315.78 & 1187.0 & 1119.0 & 6230.1 & Y    & 0.2 & 0.2\\
  1228.8202  &       40.989 &   243.97 & 2$_{20}$-2$_{11}$  &  136.16 & 1297.0 & 1271.0 & 3075.7 & N    & 0.4 & 0.4\\
  1435.0526  &       47.868 &   208.91 & 9$_{46}$-10$_{19}$ & 1340.88 &  107.3 &  100.0 &   57.8 & N    & 1.9 & 1.7\\
  1440.8088  &       48.060 &   208.08 & 7$_{26}$-6$_{33}$  &  709.61 &  775.9 &  725.2 &  845.6 & Y    & 0.9 & 0.9\\
  1542.0214  &       51.436 &   194.42 & 6$_{33}$-5$_{42}$  &  661.55 & 1129.0 & 1146.0 &  944.1 & N    & 1.2 & 1.2\\
\hline
\multicolumn{10}{l}{{\it Ortho-H$_2$O}} \\
   448.0012  &       14.944 &   669.18 & 4$_{23}$-3$_{30}$  &  300.36 & ---$^{\ast}$ &389.5 &  180.3& Y & --- & 2.2\\
   556.9384  &       18.577 &   538.30 & 1$_{10}$-1$_{01}$  &   42.37 &  486.0 &  393.9 &  348.0 & N    & 1.4 & 1.1\\
   620.7059  &       20.705 &   483.00 & 5$_{32}$-4$_{41}$  &  508.81 &  378.6 &  372.3 &  185.8 & Y    & 2.0 & 2.0\\
  1097.3879  &       36.605 &   273.19 & 3$_{12}$-3$_{03}$  &  173.37 & 1317.0 & 1382.0 & 3560.4 & Y    & 0.4 & 0.4\\
  1153.1508  &       38.465 &   259.98 & 3$_{12}$-2$_{21}$  &  173.37 & 1953.0 & 2367.0 & 5092.8 & Y    & $^{\flat}$ & $^{\flat}$ \\
  1158.3673  &       38.639 &   258.81 & 6$_{34}$-5$_{41}$  &  648.98 & 1063.0 & 1025.0 &  466.5 & Y(w) & 2.3 & 2.2\\
  1162.9117  &       38.791 &   257.79 & 3$_{21}$-3$_{12}$  &  212.16 & 1673.0 & 1635.0 & 5691.6 & Y(vw)& 0.3 & 0.3\\
  1168.3805  &       38.973 &   256.59 & 8$_{54}$-7$_{61}$  & 1255.17 &  666.7 &  668.9 &  153.7 & Y    & 4.3 & 4.4\\
  1278.2872  &       42.639 &   234.53 & 7$_{43}$-6$_{52}$  &  931.24 &  587.2 &  570.3 &  387.8 & Y    & 1.5 & 1.5\\
  1296.4552  &       43.245 &   231.25 & 8$_{27}$-7$_{34}$  &  885.60 &  907.1 &  821.1 &  804.5 & Y    & 1.1 & 1.0\\
  1307.9976  &       43.630 &   229.21 & 8$_{45}$-9$_{18}$  & 1122.71 &  135.2 &  102.3 &   42.7 & Y    & 3.2 & 2.4\\
  1322.0648  &       44.099 &   226.76 & 6$_{25}$-5$_{32}$  &  552.91 & 1647.0 & 1492.0 & 1150.9 & Y(vw)& 1.4 & 1.3\\
  1410.6491  &       47.054 &   212.53 & 5$_{23}$-5$_{14}$  &  446.51 & 1512.0 & 2989.0 & 4815.6 & Y    & 0.3 & 0.6 \\
  1574.2321  &       52.510 &   190.44 & 6$_{43}$-7$_{16}$  &  756.72 &  232.9 & ---$^{\ast}$   &  169.5 & Y    & 1.4 & ---\\
\hline 
\end{tabular}
\\

$^{\S}$: Observed line intensities (OD 123).
$^{\dag}$: {\sc SMMOL} predicted line intensities.
$^{\ddag}$: {\sc SMMOL} prediction for population inversion: Y (yes) or N (no). Y(w) indicates population inversion, but weakly.
$^{\P}$: the ratio of the observed to predicted line fluxes.
$^{\flat}$: blended with CO $J$=10--9.
$^{\ast}$: the line was out of the spectral coverage
\end{center}
\end{table*}

\begin{figure*}
\centering
\resizebox{0.9\hsize}{!}{\includegraphics*[45,298][573,769]{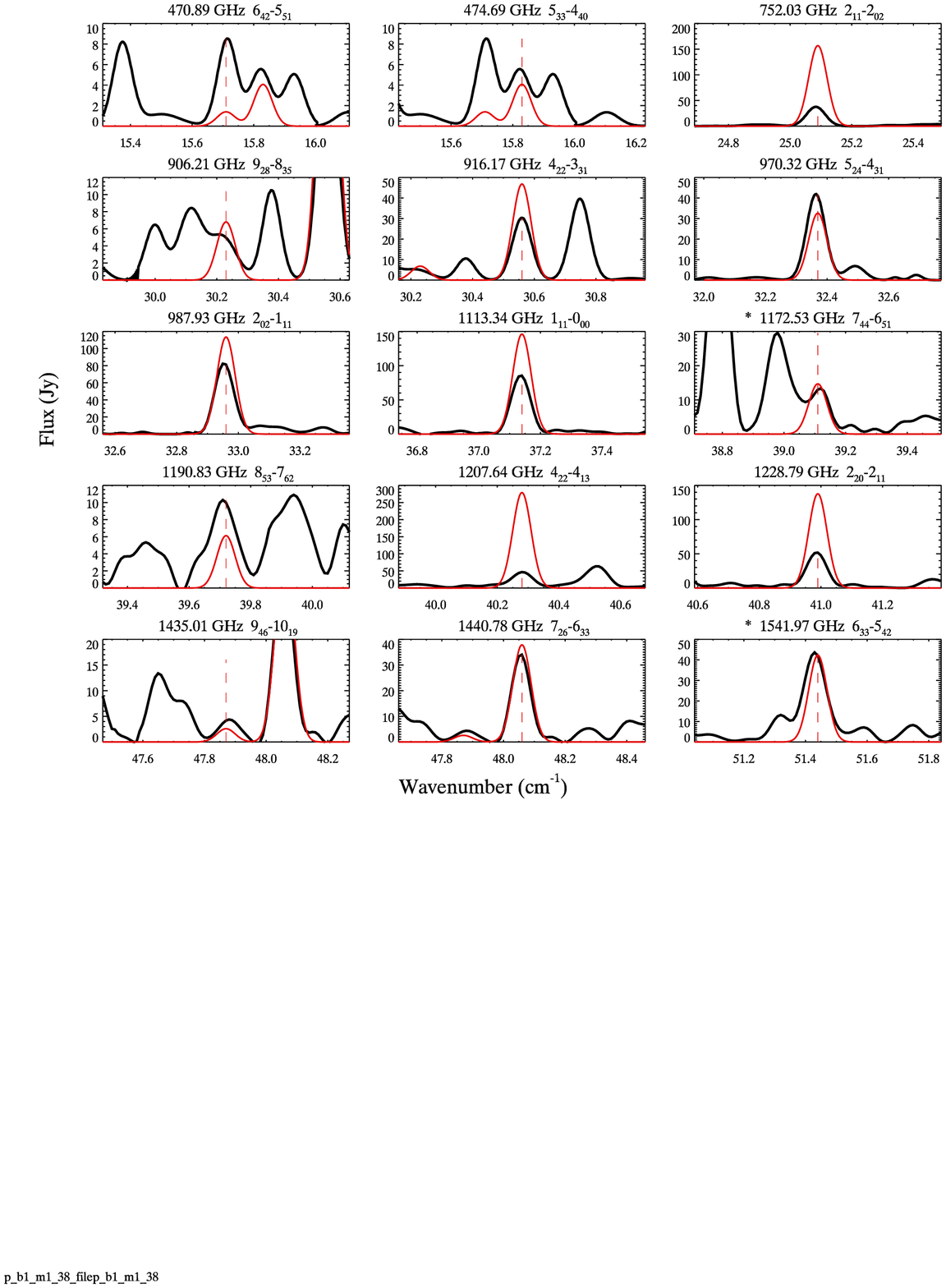}}
\caption{ {\sc smmol} fits to the para-H$_2$O lines observed by
SPIRE. The OD~123 spectra are plotted in bold black while the model
spectra are in red. The frequencies of the lines, denoted by vertical
dashes, are given at the top of the panels.
}
\label{fig-smmol-para-1}
\end{figure*}
\begin{figure*}
\centering
\resizebox{0.9\hsize}{!}{\includegraphics*[45,298][573,769]{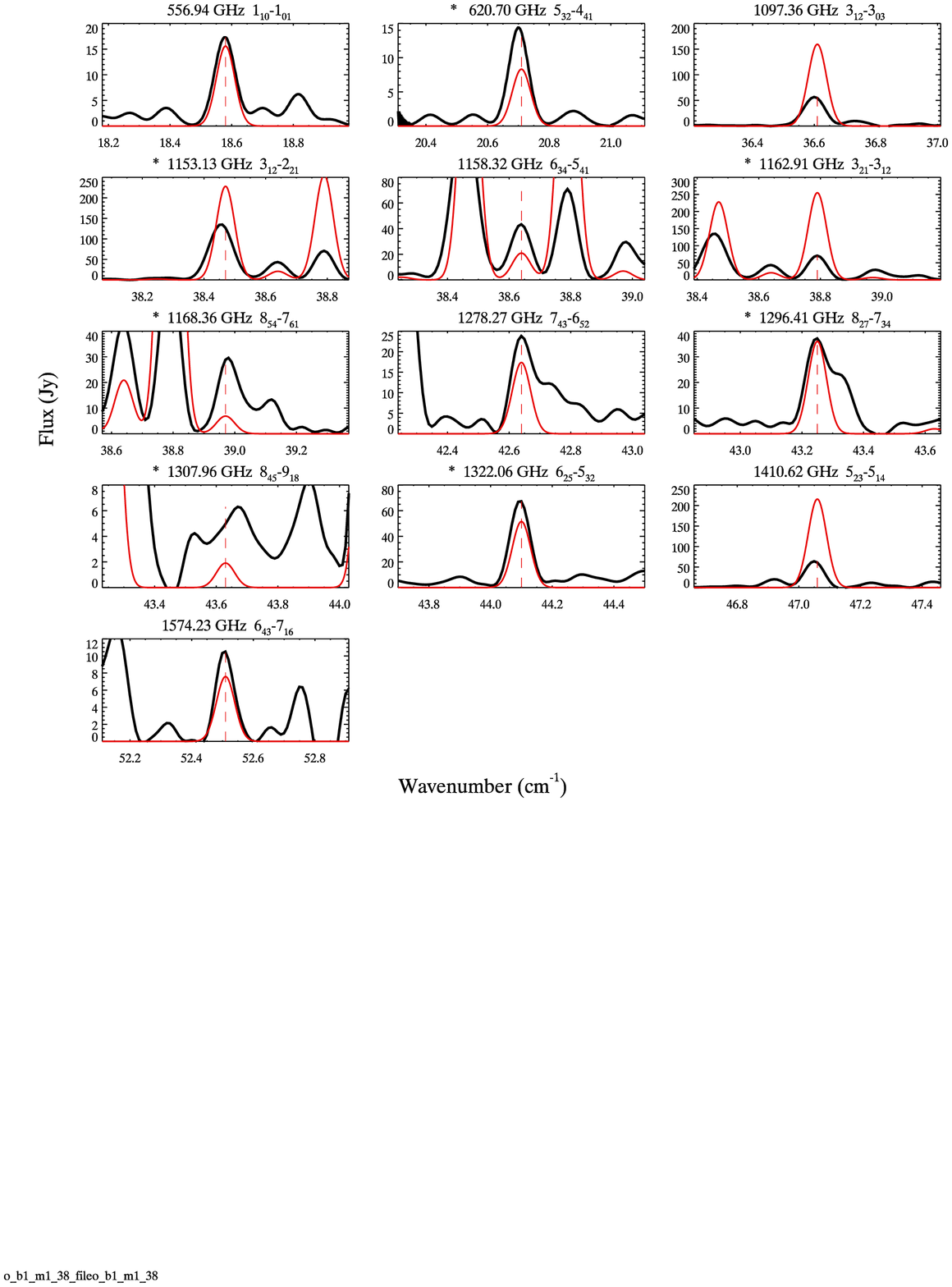}}
\caption{ {\sc smmol} fits to the ortho-H$_2$O lines observed by
SPIRE. The OD~123 spectra are plotted in black while the model
spectra are in red. The frequencies of the lines, denoted by vertical
dashes, are given at the top of the panels.
}
\label{fig-smmol-ortho-1}
\end{figure*}

The non-LTE model line intensities for $^{12}$CO, $^{13}$CO and SiO are 
plotted in the previously discussed spectral line energy distribution 
diagrams (Figs.\ref{fig-energy-CO}, \ref{fig-energy-13CO} and 
\ref{fig-energy-SiO}, respectively), whereas the modelled and observed 
H$_2$O lines are compared via spectral plots, since H$_2$O line 
intensities do not follow a simple curve in a spectral line energy 
distribution diagram. Fig\,\ref{fig-smmol-para-1} (para-H$_2$O) and 
Fig\,\ref{fig-smmol-ortho-1} (ortho-H$_2$O) show the model spectra for 25 
different H$_2$O lines, overlaid on the SPIRE observations of the lines. 
The model spectra have been convolved to the apodized instrumental 
resolution, since the observed lines are unresolved. 

We have estimated the $\chi^2$ goodness of fit, defined by $\chi^2 = \Sigma 
((O_i - P_i)^2 / \sigma_i^2 )/ n$, where $O_i$ is the SPIRE line 
intensity, $P_i$ is the {\sc smmol}-predicted line intensity and $n$ is 
the number of degrees of freedom. The uncertainty $\sigma_i$ in an 
observed line flux was adopted to be 20\,\% of the observed flux,
accounting the difference of the measured line intensities between OD 123 and 317. Table\,\ref{table-chi-sq} summarises the 
$\chi^2$ values obtained for the different molecular species. CO and SiO 
show the best fits. Table\,\ref{table-chi-sq} also summarises the observed 
to predicted line flux ratios for the different molecular species that 
were modelled.

\begin{figure*}
\centering
\resizebox{0.8\hsize}{!}{\includegraphics*{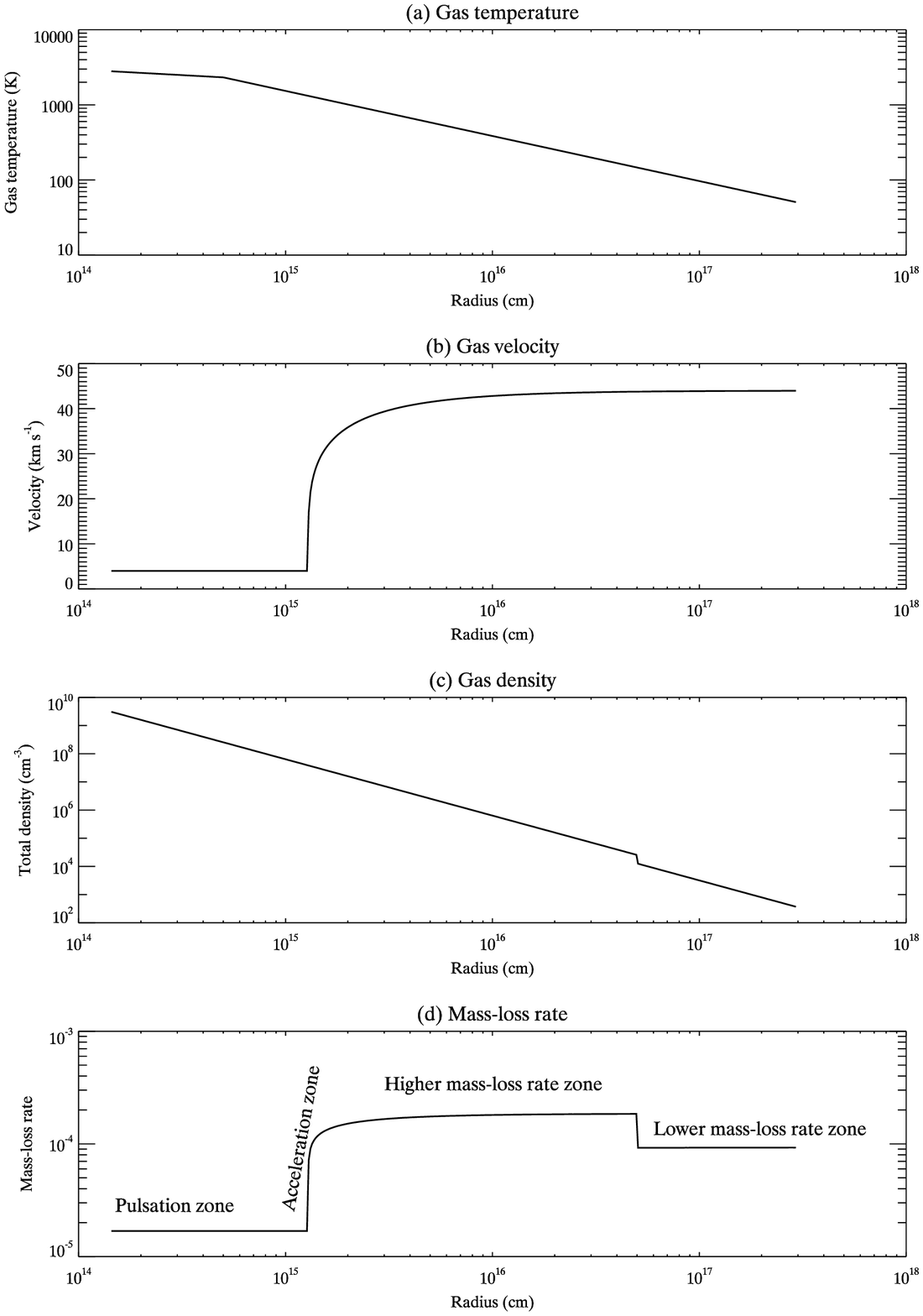}}
\caption{The wind structure required by the best-fit {\sc smmol} model,
whose parameters are listed in Table\,\ref{table-smmol-para}. 
\label{fig-structure}}
\end{figure*}

Our overall best-fitting model (see Table \ref{table-smmol-para} and 
Figure\,\ref{fig-structure}) required a gas to dust mass ratio of 267, 
with a gas acceleration zone (to a terminal velocity of 44~km~s$^{-1}$) 
located at the dust condensation radius at 9~$R_*$, beyond which the total 
mass loss rate in the region labelled as the `higher mass loss rate zone' 
in Figure\,\ref{fig-structure} is $1.85\times 
10^{-4}$~M$_{\odot}$\,yr$^{-1}$, dropping by a factor of two, to 
$9.5\times 10^{-5}$~M$_{\odot}$\,yr$^{-1}$, beyond a radius 
$R_{outflow, break} \sim 350~R_*$. 

The model required a wind with a two-component radial temperature 
profile. The higher-excitation rotational lines of water required a flat 
($\alpha \sim 0.1-0.15$) radial power-law temperature profile from the 
stellar photosphere out to (1.0--1.3)$\times10^{15}$~cm (7--9~$R_*$), 
close to the radius of the dust condensation zone suggested by our dust 
radiative transfer calculations. Beyond there, the CO lines and the 
lower-excitation water lines could be fitted with an $r^{-0.5}$ 
temperature distribution. Initial attempts were made to use a single 
radial temperature profile for the CO and H$_{2}$O models but in both 
cases too little flux was produced in high-$J$ CO transitions
and high $E_{\rm up}$ H$_2$O transitions, 
although they successfully reproduced the lower transitions. We 
detected strong vibrationally excited lines of water in our {\em Herschel} 
spectra, which require temperatures in excess of 1000\,K to be 
significantly populated by collisions. Such lines will also be optically 
thick and it is this that maintains a relatively flat temperature profile, 
because heat cannot easily escape by radiative processes. However, once 
the gas cools to about 1000~K the vibrational states are no longer 
collisionally excited, thus removing a large source of opacity and 
allowing photons to escape. At this point the radial profile adopts the 
expected $r^{-0.5}$ profile. This behaviour was first predicted by 
\citet[][Figure~2]{Goldreich:1976p27375}. This sudden cooling effect due 
to loss of opacity can trigger the formation of dust particles in the 
outflow, which up to then had been driven by radiation pressure acting 
through the opacity of vibrationally excited water. 

The modelling of our {\em Herschel} CO and H$_{2}$O line data for VY\,CMa 
(Fig.\,\ref{fig-energy-CO} for CO and Figs.\,\ref{fig-smmol-para-1} and
\ref{fig-smmol-ortho-1} for H$_2$O) 
is consistent with VY~CMa having undergone mass-loss enhancements 
in the past \citep{Richards:1998p27089, Decin:2006p27245}. 
Figure\,\ref{fig-structure} shows the mass-loss rate history of VY~CMa 
that provided the calculated molecular emission fluxes most consistent 
with our observations.  Starting from the right hand side (the older, and 
outer, part of the outflow) there was a phase of lower mass loss that 
ended 360 years ago, when the mass loss rate doubled to 
1.85$\times10^{-4}$~M${_\odot}$~yr$^{-1}$ (the `higher mass-loss rate 
zone'). The acceleration zone is 
interpreted as caused by radiative pressure on newly formed dust grains, 
followed by momentum transfer to the gas via collisions with dust grains.

The modelling of the water lines used an ortho:para ratio of 3:1 and 
included masing in the calculation of the level populations.
In general, our non-LTE models with masing included can fit the observed 
H$_2$O line intensities, but the 470.9~GHz ortho-H$_2$O line is a clear 
outlier (Table\,\ref{table-smmol-maser}) and was omitted in the 
evaluation of the $\chi^2$ and mean and median observed/predicted line 
flux ratios for H$_2$O presented in Table\,\ref{table-chi-sq}. Inspection 
of those mean and median ratios indicates that the assumption of a 3:1 
ortho:para ratio is consistent with the observations.
 The early NLTE analysis of \citet{Royer:2010p29058} obtained an H$_2$O 
ortho:para ratio of 1.3:1. We attribute the difference between the two ratios to the fact that 
\citeauthor{Royer:2010p29058}'s model was more suited to optically thin regimes, whereas {\sc smmol} is 
designed to handle large optical depths and includes maser effects.

The LTE analysis of the SiO lines clearly showed a temperature structure
with at least two components (Fig.\,\ref{fig-energy-SiO}). This was
confirmed by our non-LTE radiative transfer modelling of the CO and
H$_{2}$O lines. The inferred relatively flat radial temperature profile
out to 7--9 stellar radii, coupled with the detection of strong
vibrationally excited water lines, indicates the existence of a region
where opacity from vibrationally excited lines traps the radiation and
prevents radiative cooling of the gas, confirming the existence of the
flat inner radial temperature profiles predicted by
\citeauthor{Goldreich:1976p27375}. The modelling of SiO required a
significant abundance drop at the dust forming region, consistent with a
large fraction of SiO molecules condensing into silicate dust grains. Our
model assumes a factor of 20 reduction in the density of SiO.

\subsection{Masing and lasing transitions of water vapour}

For the water lines modelled with {\sc smmol} our non-LTE radiative 
transfer calculations allowed us to infer which ones should be undergoing 
masing action. Table\,\ref{table-smmol-maser} lists the H$_2$O transitions 
which are predicted by our best-fit {\sc smmol} model to have inverted 
energy populations and to be undergoing masing action (70\%).  For ease of 
comparison, the observed line fluxes included in the table are from the 
OD123 SPIRE FTS observations only. Our best fit model predicts six masing 
lines below 1000\,GHz and 14 lasing lines above 1000\,GHz, so the 
widespread occurrence of masing or lasing action by water molecules in the 
outflow is expected. This is consistent with the predictions of 
\citet{Deguchi:1989p28237} and \citet{Neufeld:1991ky}, who used line 
escape probability techniques, and of \citet{Doty:1997hh} and 
\citet{Yates:1997p28231}, who both used accelerated lambda iteration 
techniques to model the line radiative transfer.

Higher spectral resolution observations could test whether maser action is 
occurring for the predicted lines. Velocity-resolved line profiles have 
been already measured for three of the lines that are predicted to 
be masing. From {\em Herschel} HIFI observations of VY\,CMa, 
\citet{Harwit:2010p29900} found the H$_2$O $5_{32}$--$4_{41}$ transition 
at 620.70\,GHz (20.70\,cm$^{-1}$) to show a very narrowly peaked line 
profile, suggesting non-thermal maser emission. Ground-based APEX 
observations by \citet{Menten:2008p29901} of the para-H$_2$O 
$6_{42}$--$5_{51}$ transition at 470.889\,GHz (15.70\,cm$^{-1}$) and the 
$5_{33}$--$4_{40}$ transition at 474.680\,GHz (15.83\,cm$^{-1}$) showed 
several narrow emission peaks superposed on a broader overall emission 
profile.

 \citet{Humphreys:1997wy}
 suggested the detection SiO v=1 and v=2 J=7-6 and 8-7 maser emission from VY CMa. 
 Our {\sc smmol} models for SiO predict that $v=0$ low-$J$ transitions (up to $J$=7--6) 
 should be strongly masing, but for higher levels, population inversions are not predicted. 
 The inversions predicted for low-$J$ SiO levels are consistent with the strongly polarised 
 $v=$0 $J=$1--0 SiO emission found by \citet{McIntosh:2009dq}, indicating masing. 
 Caution is required however, as the future inclusion of vibrationally excited SiO transitions in the models could change the predictions for high-$J$ $v=$0 SiO lines.

\section{Summary}

The fluxes of eight $^{12}$CO line present in the 54--210-$\mu$m {\em 
Herschel} PACS spectrum of VY~CMa have been measured, along with the fluxes 
of more than 260 emission lines present in the SPIRE FTS 190-650-$\mu$m 
spectrum. Many lines are identified with H$_2$O transitions. Other identified 
molecules include CO, $^{13}$CO, H$_2^{18}$O, SiO, HCN, SO, SO$_2$, CS, 
H$_2$S, and NH$_3$. LTE excitation temperatures have been derived for CO, 
$^{13}$CO, HCN and SiO, while non-LTE modelling has been carried out for the 
$^{12}$CO lines in the PACS and SPIRE spectra, and for the$^{13}$CO, H$_2$O 
and SiO lines in the SPIRE FTS spectra.

Our LTE and non-LTE analyses have shown the following:
\begin{itemize}

\item Corresponding transitions of $^{12}$CO and $^{13}$CO in the SPIRE 
FTS spectrum have a flux ratio of 7.0$\pm$1.8. These lines can be fitted 
with an LTE excitation temperature of 350$\pm$90~K and a $^{12}$C/$^{13}$C 
ratio of 6.2$\pm$2.0. Our best fitting non-LTE model yields a 
$^{12}$C/$^{13}$C ratio of 5.6$\pm$1.8, much lower than previously derived 
from ground-based measurements of lower-$J$ lines but consistent with 
$^{12}$C/$^{13}$C ratios measured previously for other M supergiants.

\item More than twenty pure rotational transitions of SiO were detected in 
the SPIRE FTS spectra. The spectral line energy distribution of these 
lines clearly shows two SiO temperature components: one at $\sim$1000\,K 
and and the other at $\sim$200\,K. The first appears to correspond to the 
region between the photosphere and the dust condensation radius, and the 
latter to the region of the circumstellar envelope beyond the dust 
condensation radius.

\item HCN was detected in the {\em Herschel} spectra, as already known 
from ground-based millimetre wavelength observations. The formation 
process for HCN in an oxygen-rich circumstellar envelopes is still uncertain, 
with the two main candidates being photon-mediated processes in the outer 
circumstellar envelope and shock chemistry in the stellar atmosphere and 
inner wind. The low excitation temperature (250\,K) that we derive for 
HCN, lower even than that of CO, favours the first of these mechanisms.

\item Our best-fit non-LTE model for $^{12}$CO, $^{13}$CO and H$_2$O 
yields an overall mass-loss rate of 
$1.85\times10^{-4}$\,$M_{\odot}$\,yr$^{-1}$ between radii of 9~R$_*$ and 
350~R$_*$, dropping by a factor of two beyond that. It requires a flat 
temperature profile below the dust condensation radius, consistent with 
the H$_2$O opacity preventing coolant lines from escaping, while beyond 
that point the gas temperature follows an $r^{-0.5}$ dependence \citep[as 
predicted by][]{Goldreich:1976p27375}, consistent with the lines having 
become optically thin.

\item The NLTE H$_2$O line modelling predicts that many of the lines in 
the SPIRE frequency band should be masing. The observed fluxes alone are not 
sufficient to confirm masing in the lines, although previously published 
high spectral resolution observations have indicated maser action in at 
least one of the lines.

\end{itemize}

\section{acknowledgments}
SPIRE has been developed by a consortium of institutes led by Cardiff 
University (UK) and including Univ. Lethbridge (Canada); NAOC (China); 
CEA, LAM (France); IFSI, Univ. Padua (Italy); IAC (Spain); Stockholm 
Observatory (Sweden); Imperial College London, RAL, UCL-MSSL, UKATC, Univ. 
Sussex (UK); and Caltech, JPL, NHSC, Univ. Colorado (USA). This 
development has been supported by national funding agencies: CSA (Canada); 
NAOC (China); CEA, CNES, CNRS (France); ASI (Italy); MCINN (Spain); SNSB 
(Sweden); STFC and UKSA (UK); and NASA (USA).
PACS has been developed by a consortium of institutes led by MPE (Germany) 
and including UVIE (Austria); KU Leuven, CSL, IMEC (Belgium); CEA, LAM 
(France); MPIA (Germany); INAF-IFSI/OAA/OAP/OAT, LENS, SISSA (Italy); IAC 
(Spain). This development has been supported by the funding agencies BMVIT 
(Austria), ESA-PRODEX (Belgium), CEA/CNES (France), DLR (Germany), 
ASI/INAF (Italy), and CICYT/MCYT (Spain).
PvH acknowledges support from the Belgian Science Policy office through the ESA PRODEX program.

\bibliography{vycma}

\begin{thebibliography}{}

\bibitem[\protect\citeauthoryear{Barber, Tennyson, Harris \& Tolchenov}{Barber
  et~al.}{2006}]{Barber:2006p29886}
Barber R.~J.,  Tennyson J.,  Harris G.~J.,    Tolchenov R.~N.,  2006, \mnras,
  368, 1087

\bibitem[\protect\citeauthoryear{Barlow, Nguyen-Q-Rieu, Truong-Bach \& et
  al.}{Barlow et~al.}{1996}]{Barlow:1996tj}
Barlow M.~J.,  Nguyen-Q-Rieu Truong-Bach   et al. 1996, A{\&}A, 315, L241

\bibitem[\protect\citeauthoryear{Beichman, Neugebauer, Habing, Clegg \&
  Chester}{Beichman et~al.}{1988}]{Beichman:1988p29907}
Beichman C.~A.,  Neugebauer G.,  Habing H.~J.,  Clegg P.~E.,    Chester T.~J.,
  1988, Infrared astronomical satellite (IRAS) catalogs and atlases. Volume 1,
  1

\bibitem[\protect\citeauthoryear{{Biermann} \& {Harwit}}{{Biermann} \&
  {Harwit}}{1980}]{biermann80}
{Biermann} P.,  {Harwit} M.,  1980, \apjl, 241, L105

\bibitem[\protect\citeauthoryear{Charnley, Tielens \& Kress}{Charnley
  et~al.}{1995}]{Charnley:1995wo}
Charnley S.~B.,  Tielens A. G. G.~M.,    Kress M.~E.,  1995, MNRAS, 274, L53

\bibitem[\protect\citeauthoryear{Choi, Hirota, Honma \& et al.}{Choi
  et~al.}{2008}]{Choi:2008p27073}
Choi Y.~K.,  Hirota T.,  Honma M.,    et al. 2008, PASJ, 60, 1007

\bibitem[\protect\citeauthoryear{Clegg, Ade, Armand \& et al.}{Clegg
  et~al.}{1996}]{Clegg:1996p26437}
Clegg P.~E.,  Ade P. A.~R.,  Armand C.,    et al. 1996, A{\&}A, 315, L38

\bibitem[\protect\citeauthoryear{{Danchi}, {Bester}, {Degiacomi}, {Greenhill}
  \& {Townes}}{{Danchi} et~al.}{1994}]{Danchi:1994aj}
{Danchi} W.~C.,  {Bester} M.,  {Degiacomi} C.~G.,  {Greenhill} L.~J.,
  {Townes} C.~H.,  1994, \aj, 107, 1469

\bibitem[\protect\citeauthoryear{Dayou \& Balan{\c c}a}{Dayou \& Balan{\c
  c}a}{2006}]{Dayou:2006ey}
Dayou F.,  Balan{\c c}a C.,  2006, A{\&}A, 459, 297

\bibitem[\protect\citeauthoryear{De~Beck, Decin, De~Koter, Justtanont,
  Verhoelst, Kemper \& Menten}{De~Beck et~al.}{2010}]{DeBeck:2010fg}
De~Beck E.,  Decin L.,  De~Koter A.,  Justtanont K.,  Verhoelst T.,  Kemper F.,
     Menten K.~M.,  2010, A{\&}A, 523, A18

\bibitem[\protect\citeauthoryear{de Graauw, Helmich, Phillips \& et
  al.}{de~Graauw et~al.}{2010}]{deGraauw:2010p29902}
de Graauw T.,  Helmich F.~P.,  Phillips T.~G.,    et al. 2010, \aap, 518, L6

\bibitem[\protect\citeauthoryear{Decin, Hony, Koter, Justtanont, Tielens \&
  Waters}{Decin et~al.}{2006}]{Decin:2006p27245}
Decin L.,  Hony S.,  Koter A.~D.,  Justtanont K.,  Tielens A. G. G.~M.,
  Waters L. B. F.~M.,  2006, A{\&}A, 456, 549

\bibitem[\protect\citeauthoryear{Deguchi \& Goldsmith}{Deguchi \&
  Goldsmith}{1985}]{Deguchi:1985kb}
Deguchi S.,  Goldsmith P.~F.,  1985, Nature, 317, 336

\bibitem[\protect\citeauthoryear{Deguchi \& Nguyen-Q-Rieu}{Deguchi \&
  Nguyen-Q-Rieu}{1990}]{Deguchi:1990p27374}
Deguchi S.,  Nguyen-Q-Rieu 1990, \apj, 360, L27

\bibitem[\protect\citeauthoryear{Deguchi \& Watson}{Deguchi \&
  Watson}{1989}]{Deguchi:1989p28237}
Deguchi S.,  Watson W.~D.,  1989, \apj, 340, L17

\bibitem[\protect\citeauthoryear{Doty \& Neufeld}{Doty \&
  Neufeld}{1997}]{Doty:1997hh}
Doty S.~D.,  Neufeld D.~A.,  1997, \apj, 489, 122

\bibitem[\protect\citeauthoryear{Draine \& Lee}{Draine \&
  Lee}{1984}]{Draine:1984jw}
Draine B.~T.,  Lee H.~M.,  1984, \apj, 285, 89

\bibitem[\protect\citeauthoryear{Duari, Cherchneff \& Willacy}{Duari
  et~al.}{1999}]{Duari:1999p27058}
Duari D.,  Cherchneff I.,    Willacy K.,  1999, A{\&}A, 341, L47

\bibitem[\protect\citeauthoryear{Efstathiou \& Rowan-Robinson}{Efstathiou \&
  Rowan-Robinson}{1990}]{Efstathiou:1990p27044}
Efstathiou A.,  Rowan-Robinson M.,  1990, \mnras, 245, 275

\bibitem[\protect\citeauthoryear{Ercolano, Barlow \& Storey}{Ercolano
  et~al.}{2005}]{Ercolano:2005p29908}
Ercolano B.,  Barlow M.~J.,    Storey P.~J.,  2005, \mnras, 362, 1038

\bibitem[\protect\citeauthoryear{Faure \& Josselin}{Faure \&
  Josselin}{2008}]{Faure:2008p28232}
Faure A.,  Josselin E.,  2008, A{\&}A, 492, 257

\bibitem[\protect\citeauthoryear{Fulton, Baluteau, Bendo \& et al.}{Fulton
  et~al.}{2010}]{Fulton:2010gt}
Fulton T.~R.,  Baluteau J.-P.,  Bendo G.,    et al. 2010, in Oschmann Jr J.~M.,
   Clampin M.~C.,   MacEwen H.~A.,  eds, SPIE Astronomical Telescopes and
  Instrumentation: Observational Frontiers of Astronomy for the New Decade {The
  data processing pipelines for the Herschel/SPIRE imaging Fourier transform
  spectrometer}.
SPIE, pp 773134--773134--12

\bibitem[\protect\citeauthoryear{Fulton, Naylor, Baluteau, Griffin,
  Davis-Imhof, Swinyard, Lim, Surace, Clements, Panuzzo, Gastaud, Polehampton,
  Guest, Lu, Schwartz \& Xu}{Fulton et~al.}{2008}]{Fulton:2008bt}
Fulton T.~R.,  Naylor D.~A.,  Baluteau J.-P.,  Griffin M.,  Davis-Imhof P.,
  Swinyard B.~M.,  Lim T.~L.,  Surace C.,  Clements D.,  Panuzzo P.,  Gastaud
  R.,  Polehampton E.,  Guest S.,  Lu N.,  Schwartz A.,    Xu K.,  2008, in
  Oschmann Jr J.~M.,  de Graauw M. W.~M.,   MacEwen H.~A.,  eds, Astronomical
  Telescopes and Instrumentation: Synergies Between Ground and Space {The data
  processing pipeline for the Herschel/SPIRE imaging Fourier Transform
  Spectrometer}.
SPIE, pp 70102T--70102T--12

\bibitem[\protect\citeauthoryear{Goldreich \& Scoville}{Goldreich \&
  Scoville}{1976}]{Goldreich:1976p27375}
Goldreich P.,  Scoville N.,  1976, \apj, 205, 144

\bibitem[\protect\citeauthoryear{Goldsmith \& Langer}{Goldsmith \&
  Langer}{1999}]{Goldsmith:1999ia}
Goldsmith P.~F.,  Langer W.~D.,  1999, \apj, 517, 209

\bibitem[\protect\citeauthoryear{Green, Maluendes \& McLean}{Green
  et~al.}{1993}]{Green:1993p28623}
Green S.,  Maluendes S.,    McLean A.~D.,  1993, \apjs, 85, 181

\bibitem[\protect\citeauthoryear{Griffin \& et al.}{Griffin \&
  et~al.}{2010}]{Griffin:2010p29303}
Griffin M.~J.,  et al. A.~A.,  2010, \aap, 518, L3

\bibitem[\protect\citeauthoryear{Groenewegen, Waelkens, Barlow \& et
  al.}{Groenewegen et~al.}{2011}]{Groenewegen:2011hi}
Groenewegen M. A.~T.,  Waelkens C.,  Barlow M.~J.,    et al. 2011, \aap, 526,
  162

\bibitem[\protect\citeauthoryear{Harwit, Houde, Sonnentrucker \& et al.}{Harwit
  et~al.}{2010}]{Harwit:2010p29900}
Harwit M.,  Houde M.,  Sonnentrucker P.,    et al. 2010, \aap, 521, L51

\bibitem[\protect\citeauthoryear{Humphreys, Gray, Yates \& Field}{Humphreys
  et~al.}{1997}]{Humphreys:1997wy}
Humphreys E. M.~L.,  Gray M.~D.,  Yates J.~A.,    Field D.,  1997, Monthly
  Notices of the Royal Astronomical Society, 287, 663

\bibitem[\protect\citeauthoryear{Humphreys, Helton \& Jones}{Humphreys
  et~al.}{2007}]{Humphreys:2007p26971}
Humphreys R.~M.,  Helton L.~A.,    Jones T.~J.,  2007, \aj, 133, 2716

\bibitem[\protect\citeauthoryear{Ireland, Scholz \& Wood}{Ireland
  et~al.}{2011}]{Ireland:2011ij}
Ireland M.~J.,  Scholz M.,    Wood P.~R.,  2011, MNRAS, 418, 114

\bibitem[\protect\citeauthoryear{{Jewell}, {Snyder} \& {Schenewerk}}{{Jewell}
  et~al.}{1986}]{Jewell:86}
{Jewell} P.~R.,  {Snyder} L.~E.,    {Schenewerk} M.~S.,  1986, \nat, 323, 311

\bibitem[\protect\citeauthoryear{Justtanont, Barlow, Tielens, Hollenbach,
  Latter, Liu, Sylvester, Cox \& Rieu}{Justtanont
  et~al.}{2000}]{Justtanont:2000p27249}
Justtanont K.,  Barlow M.~J.,  Tielens A. G. G.~M.,  Hollenbach D.,  Latter
  W.~B.,  Liu X.-W.,  Sylvester R.~J.,  Cox P.,    Rieu N.-Q.,  2000, A{\&}A,
  360, 1117

\bibitem[\protect\citeauthoryear{Justtanont, Khouri, Maercker \& et
  al.}{Justtanont et~al.}{2011}]{Justtanont:2011wi}
Justtanont K.,  Khouri T.,  Maercker M.,    et al. 2011, A{\&}A

\bibitem[\protect\citeauthoryear{Kaminski, Gottlieb, Menten, Patel, Young,
  Br{\"u}nken, M{\"u}ller, McCarthy, Winters \& Decin}{Kaminski
  et~al.}{2013}]{Kaminski:2013gk}
Kaminski T.,  Gottlieb C.~A.,  Menten K.~M.,  Patel N.~A.,  Young K.~H.,
  Br{\"u}nken S.,  M{\"u}ller H. S.~P.,  McCarthy M.~C.,  Winters J.~M.,
  Decin L.,  2013, A\&A, 551, A113

\bibitem[\protect\citeauthoryear{Kaufman \& Neufeld}{Kaufman \&
  Neufeld}{1996}]{Kaufman:1996p26003}
Kaufman M.~J.,  Neufeld D.~A.,  1996, \apj, 456, 611

\bibitem[\protect\citeauthoryear{Kemper, Stark, Justtanont, De~Koter, Tielens,
  Waters, Cami \& Dijkstra}{Kemper et~al.}{2003}]{Kemper:2003kt}
Kemper F.,  Stark R.,  Justtanont K.,  De~Koter A.,  Tielens A. G. G.~M.,
  Waters L. B. F.~M.,  Cami J.,    Dijkstra C.,  2003, \aap, 407, 609

\bibitem[\protect\citeauthoryear{Langhoff \& Bauschlicher}{Langhoff \&
  Bauschlicher}{1993}]{Langhoff:1993p26503}
Langhoff S.~R.,  Bauschlicher C.~W.,  1993, Chemical Physics Letters, 211, 305

\bibitem[\protect\citeauthoryear{Lerate, Yates, Barlow, Viti \&
  Swinyard}{Lerate et~al.}{2010}]{Lerate:2010cj}
Lerate M.~R.,  Yates J.~A.,  Barlow M.~J.,  Viti S.,    Swinyard B.~M.,  2010,
  \mnras, 406, 2445

\bibitem[\protect\citeauthoryear{McIntosh \& Rislow}{McIntosh \&
  Rislow}{2009}]{McIntosh:2009dq}
McIntosh G.~C.,  Rislow B.,  2009, The Astrophysical Journal, 692, 154

\bibitem[\protect\citeauthoryear{Mamon, Glassgold \& Huggins}{Mamon
  et~al.}{1988}]{Mamon:1988br}
Mamon G.~A.,  Glassgold A.~E.,    Huggins P.~J.,  1988, \apj, 328, 797

\bibitem[\protect\citeauthoryear{{Mathis}, {Rumpl} \& {Nordsieck}}{{Mathis}
  et~al.}{1977}]{mathis77}
{Mathis} J.~S.,  {Rumpl} W.,    {Nordsieck} K.~H.,  1977, \apj, 217, 425

\bibitem[\protect\citeauthoryear{Matsuura, Yamamura, Cami, Onaka \&
  Murakami}{Matsuura et~al.}{2002}]{Matsuura:2002p25132}
Matsuura M.,  Yamamura I.,  Cami J.,  Onaka T.,    Murakami H.,  2002, A{\&}A,
  383, 972

\bibitem[\protect\citeauthoryear{Menten, Lundgren, Belloche, Thorwirth \&
  Reid}{Menten et~al.}{2008}]{Menten:2008p29901}
Menten K.~M.,  Lundgren A.,  Belloche A.,  Thorwirth S.,    Reid M.~J.,  2008,
  \aap, 477, 185

\bibitem[\protect\citeauthoryear{Menten, Philipp, G{\"u}sten, Alcolea,
  Polehampton \& Br{\"u}nken}{Menten et~al.}{2006}]{Menten:2006p27227}
Menten K.~M.,  Philipp S.~D.,  G{\"u}sten R.,  Alcolea J.,  Polehampton E.~T.,
    Br{\"u}nken S.,  2006, A{\&}A, 454, L107

\bibitem[\protect\citeauthoryear{Menten, Wyrowski, Alcolea \& et al.}{Menten
  et~al.}{2010}]{Menten:2010p29072}
Menten K.~M.,  Wyrowski F.,  Alcolea J.,    et al. 2010, \aap, 521, L7

\bibitem[\protect\citeauthoryear{Menten \& Young}{Menten \&
  Young}{1995}]{Menten:1995gt}
Menten K.~M.,  Young K.,  1995, The Astrophysical Journal, 450, L67

\bibitem[\protect\citeauthoryear{Milam, Woolf \& Ziurys}{Milam
  et~al.}{2009}]{Milam:2009p15019}
Milam S.~N.,  Woolf N.~J.,    Ziurys L.~M.,  2009, \apj, 690, 837

\bibitem[\protect\citeauthoryear{Monnier, Tuthill, Lopez, Cruzalebes, Danchi \&
  Haniff}{Monnier et~al.}{1999}]{Monnier:1999p27021}
Monnier J.~D.,  Tuthill P.~G.,  Lopez B.,  Cruzalebes P.,  Danchi W.~C.,
  Haniff C.~A.,  1999, \apj, 512, 351

\bibitem[\protect\citeauthoryear{M{\"u}ller, Schl{\"o}der, Stutzki \&
  Winnewisser}{M{\"u}ller et~al.}{2005}]{Muller:2005p29905}
M{\"u}ller H. S.~P.,  Schl{\"o}der F.,  Stutzki J.,    Winnewisser G.,  2005,
  Journal of Molecular Structure, 742, 215

\bibitem[\protect\citeauthoryear{Muller, Dinh-V-Trung, Lim, Hirano, Muthu \&
  Kwok}{Muller et~al.}{2007}]{Muller:2007p26972}
Muller S.,  Dinh-V-Trung Lim J.,  Hirano N.,  Muthu C.,    Kwok S.,  2007,
  \apj, 656, 1109

\bibitem[\protect\citeauthoryear{Naylor \& Tahic}{Naylor \&
  Tahic}{2007}]{Naylor:2007kl}
Naylor D.~A.,  Tahic M.~K.,  2007, Journal of the Optical Society of America A,
  24, 3644

\bibitem[\protect\citeauthoryear{Nercessian, Omont, Benayoun \&
  Guilloteau}{Nercessian et~al.}{1989}]{Nercessian:1989p27097}
Nercessian E.,  Omont A.,  Benayoun J.~J.,    Guilloteau S.,  1989, \aap, 210,
  225

\bibitem[\protect\citeauthoryear{Neufeld, Feuchtgruber, Harwit \&
  Melnick}{Neufeld et~al.}{1999}]{Neufeld:1999p27088}
Neufeld D.~A.,  Feuchtgruber H.,  Harwit M.,    Melnick G.~J.,  1999, \apj,
  517, L147

\bibitem[\protect\citeauthoryear{Neufeld \& Melnick}{Neufeld \&
  Melnick}{1991}]{Neufeld:1991ky}
Neufeld D.~A.,  Melnick G.~J.,  1991, \apj, 368, 215

\bibitem[\protect\citeauthoryear{Ott}{Ott}{2010}]{Ott:2010tla}
Ott S.,  2010, Astronomical Society of the Pacific Conference Series, 434, 139

\bibitem[\protect\citeauthoryear{Pickett}{Pickett}{1991}]{Pickett:1991p29904}
Pickett H.~M.,  1991, Journal of Molecular Spectroscopy, 148, 371

\bibitem[\protect\citeauthoryear{Pickett, Poynter, Cohen, Delitsky, Pearson \&
  M{\"u}ller}{Pickett et~al.}{1998}]{Pickett:1998jh}
Pickett H.~M.,  Poynter R.~L.,  Cohen E.~A.,  Delitsky M.~L.,  Pearson J.~C.,
   M{\"u}ller H. S.~P.,  1998, Journal of Quantitative Spectroscopy and
  Radiative Transfer, 60, 883

\bibitem[\protect\citeauthoryear{Pilbratt, Riedinger, Passvogel \& et
  al.}{Pilbratt et~al.}{2010}]{Pilbratt:2010p29312}
Pilbratt G.~L.,  Riedinger J.~R.,  Passvogel T.,    et al. 2010, \aap, 518, L1

\bibitem[\protect\citeauthoryear{Poglitsch, Waelkens, Geis \& et al.}{Poglitsch
  et~al.}{2010}]{Poglitsch:2010p28964}
Poglitsch A.,  Waelkens C.,  Geis N.,    et al. 2010, \aap, 518, L2

\bibitem[\protect\citeauthoryear{Polehampton, Menten, van~der Tak \&
  White}{Polehampton et~al.}{2010}]{Polehampton:2010p28352}
Polehampton E.~T.,  Menten K.~M.,  van~der Tak F. F.~S.,    White G.~J.,  2010,
  A{\&}A, 510, 80

\bibitem[\protect\citeauthoryear{Rawlings \& Yates}{Rawlings \&
  Yates}{2001}]{Rawlings:2001p28230}
Rawlings J. M.~C.,  Yates J.~A.,  2001, \mnras, 326, 1423

\bibitem[\protect\citeauthoryear{Richards, Yates \& Cohen}{Richards
  et~al.}{1998}]{Richards:1998p27089}
Richards A. M.~S.,  Yates J.~A.,    Cohen R.~J.,  1998, \mnras, 299, 319

\bibitem[\protect\citeauthoryear{Rothman, Gordon, Barbe \& et al.}{Rothman
  et~al.}{2009}]{Rothman:2009p28626}
Rothman L.~S.,  Gordon I.~E.,  Barbe A.,    et al. 2009, Journal of
  Quantitative Spectroscopy and Radiative Transfer, 110, 533

\bibitem[\protect\citeauthoryear{Royer, Decin, Wesson \& et al.}{Royer
  et~al.}{2010}]{Royer:2010p29058}
Royer P.,  Decin L.,  Wesson R.,    et al. 2010, \aap, 518, L145

\bibitem[\protect\citeauthoryear{Rybicki \& Hummer}{Rybicki \&
  Hummer}{1991}]{Rybicki:1991p28622}
Rybicki G.~B.,  Hummer D.~G.,  1991, \aap, 245, 171

\bibitem[\protect\citeauthoryear{Sauval \& Tatum}{Sauval \&
  Tatum}{1984}]{Sauval:1984p26502}
Sauval A.~J.,  Tatum J.~B.,  1984, \apjs, 56, 193

\bibitem[\protect\citeauthoryear{Scharmer \& Carlsson}{Scharmer \&
  Carlsson}{1985}]{Scharmer:1985p28621}
Scharmer G.~B.,  Carlsson M.,  1985, Journal of Computational Physics, 59, 56

\bibitem[\protect\citeauthoryear{Sidaner \& {Le Bertre}}{Sidaner \& {Le
  Bertre}}{1996}]{LeSidaner:1996p8500}
Sidaner P.~L.,  {Le Bertre} T.,  1996, A{\&}A, 314, 896

\bibitem[\protect\citeauthoryear{Sloan, Kraemer, Price \& Shipman}{Sloan
  et~al.}{2003}]{Sloan:2003p28681}
Sloan G.~C.,  Kraemer K.~E.,  Price S.~D.,    Shipman R.~F.,  2003, \apjs, 147,
  379

\bibitem[\protect\citeauthoryear{Swinyard, Ade, Baluteau \& et al.}{Swinyard
  et~al.}{2010}]{Swinyard:2010p29297}
Swinyard B.~M.,  Ade P.,  Baluteau J.-P.,    et al. 2010, \aap, 518, L4

\bibitem[\protect\citeauthoryear{Tenenbaum, Dodd, Milam, Woolf \&
  Ziurys}{Tenenbaum et~al.}{2010}]{Tenenbaum:2010gu}
Tenenbaum E.~D.,  Dodd J.~L.,  Milam S.~N.,  Woolf N.~J.,    Ziurys L.~M.,
  2010, \apjs, 190, 348

\bibitem[\protect\citeauthoryear{Tsuji, Ohnaka, Aoki \& Yamamura}{Tsuji
  et~al.}{1997}]{TSUJI:1997ws}
Tsuji T.,  Ohnaka K.,  Aoki W.,    Yamamura I.,  1997, A{\&}A, 320, L1

\bibitem[\protect\citeauthoryear{Wesson, Cernicharo, Barlow \& et al.}{Wesson
  et~al.}{2010}]{Wesson:2010p29903}
Wesson R.,  Cernicharo J.,  Barlow M.~J.,    et al. 2010, \aap, 518, L144

\bibitem[\protect\citeauthoryear{Willacy \& Millar}{Willacy \&
  Millar}{1997}]{Willacy:1997p27138}
Willacy K.,  Millar T.~J.,  1997, A{\&}A, 324, 237

\bibitem[\protect\citeauthoryear{Yang, Stancil, Balakrishnan \& Forrey}{Yang
  et~al.}{2010}]{Yang:2010bb}
Yang B.,  Stancil P.~C.,  Balakrishnan N.,    Forrey R.~C.,  2010, \apj, 718,
  1062

\bibitem[\protect\citeauthoryear{Yates, Field \& Gray}{Yates
  et~al.}{1997}]{Yates:1997p28231}
Yates J.~A.,  Field D.,    Gray M.~D.,  1997, \mnras, 285, 303

\bibitem[\protect\citeauthoryear{Ziurys, Milam, Apponi \& Woolf}{Ziurys
  et~al.}{2007}]{Ziurys:2007p27081}
Ziurys L.~M.,  Milam S.~N.,  Apponi A.~J.,    Woolf N.~J.,  2007, Nature, 447,
  1094

\bibitem[\protect\citeauthoryear{Ziurys, Tenenbaum, Pulliam, Woolf \&
  Milam}{Ziurys et~al.}{2009}]{Ziurys:2009p27100}
Ziurys L.~M.,  Tenenbaum E.~D.,  Pulliam R.~L.,  Woolf N.~J.,    Milam S.~N.,
  2009, \apj, 695, 1604

\end{thebibliography}

\bibliographystyle{mn2e}

\appendix

\section{Line fluxes and line identifications}

  \begin{table*}
  {\footnotesize
  \centering
  \caption{VY~CMa: CO Line fluxes  in the PACS spectra
  \label{linelist-2}}
  \begin{tabular}{r@{.}l r@{.}l r@{$\pm$}l  rrrr}
  \hline
  \multicolumn{8}{l}{PACS} & \multicolumn{2}{l}{ID} \\
  \multicolumn{2}{l}{$\bar{\lambda}_{obs}$ ($\mu$m)}  & 
  \multicolumn{2}{l}{$\bar{\nu}_{obs}$ (cm$^{-1}$)}  & 
  \multicolumn{2}{l}{F($\times$10$^{-18}$Wm$^{-2}$)} &
  \multicolumn{1}{l}{$\bar{\nu}_{o}$     (cm$^{-1}$)}  & 
  \multicolumn{1}{l}{$\bar{\nu}_{o}$     (GHz)}  & 
  \multicolumn{1}{l}{$\bar{\lambda}_{o}$ ($\mu$m)}  & 
 Transition  \\
  \hline
186&02   &  53&758  &  1479& ~69  &  53.764     &    1611.78   &  186.00  &  CO 14-13 \\
173&65   &  57&587  &  1495& ~57  &  57.593     &    1726.59   &  173.63  &  CO 15-14 \\
162&83   &  61&412  &  1402& 153  &  61.421     &    1841.33   &  162.81  &  CO 16-15 \\
153&28   &  65&239  &  1740& ~75  &  65.246     &    1956.00   &  153.27  &  CO 17-16 \\
144&79   &  69&065  &  1755& 182  &  69.068     &    2070.60   &  144.78  &  CO 18-17 \\
137&23   &  72&870  &  1273& 122  &  72.888     &    2185.12   &  137.20  &  CO 19-18 \\
124&19   &  80&520  &   972& 188  &  80.560     &    2413.90   &  124.19  &  CO 21-20 \\
118&60   &  84&317  &  1441& 164  &  84.331     &    2528.15   &  118.58  &  CO 22-21 \\
  \hline
  \end{tabular}
  }
  \end{table*}




  \begin{table*}
  {\footnotesize
  \centering
  \caption{VY CMa line fluxes ($F$) and identifications (IDs) in the SPIRE FTS SLW spectra.
  An ID row beginning with a `+' indicates that  the line is blended with that on the row above.
  $\nu_0$ (cm$^{-1}$) and $\lambda_0$ ($\mu$m) are vacuum wavenumbers and wavelengths, respectively.
  The quantum numbers given in the `transition' column are $J$ except
  where stated.  H$_2$O transitions are given in the format $J^{\prime}$$_{K_a^\prime}$$_{K_c^{\prime}}$-$J_{K_a}$$_{K_c}$.
  The listed uncertainties for the line fluxes are those from the line fitting, and absolute calibration uncertainties are not included.}
  \label{linelist-1}
\begin{tabular}{lr lr rrl ll}
  \hline
  \multicolumn{2}{l}{OD123} & \multicolumn{2}{l}{OD317} & \multicolumn{4}{l}{ID}\\\
  $\bar{\nu}_{obs}$ (cm$^{-1}$)  & F($\times$10$^{-18}$Wm$^{-2}$) & 
  $\bar{\nu}_{obs}$ (cm$^{-1}$)  & F($\times$10$^{-18}$Wm$^{-2}$) & 
  $\nu_0$ (cm$^{-1}$) & $\nu_0$ (GHz) & $\lambda_0$ ($\mu$m)  &
  Species & Transition  \\
  \hline
           &                    & 14.931 $\pm$ 0.003 &  389.5 $\pm$  32.2 &  14.944 &    448.00 & 669.18 & o-H$_2$O      & 4$_{23}$-3$_{30}$                                 \\
    15.374 &  210.3 $\pm$   8.1 & 15.371 $\pm$ 0.001 &  190.0 $\pm$  67.8 &  15.379 &    461.05 & 650.24 & CO            & 4-3                                               \\
    15.501 &   60.2 $\pm$   9.0 & 15.538 $\pm$ 0.005 &   80.0 $\pm$   4.8 &  15.489 &    464.35 & 645.62 & p-H$_2$O?     & $\nu_1$=1,5$_{33}$-3$_{44}$                       \\
           &                    &                    &                    & +15.501 &    464.71 & 645.11 & SiS?          & v=3, 26-25                                        \\
           &                    &                    &                    & +15.505 &    464.83 & 644.96 & p-H$_2$O?     & $\nu_1$=1,10$_{37}$-11$_{2\,10}$                  \\
    15.714 &  221.8 $\pm$   4.9 & 15.707 $\pm$ 0.001 &  135.7 $\pm$   2.4 &  15.707 &    470.88 & 636.66 & p-H$_2$O      & 6$_{42}$-5$_{51}$                                 \\
\hline
  \end{tabular}
  }
  \end{table*}
  \addtocounter{table}{-1}
  \begin{table*}
  {\footnotesize
  \centering
  \caption{-- continued.}
  \begin{tabular}{lr lr rrl ll}
  \hline
  \multicolumn{2}{l}{OD123} & \multicolumn{2}{l}{OD317} & \multicolumn{4}{l}{ID}\\\
  $\bar{\nu}_{obs}$ (cm$^{-1}$)  & F($\times$10$^{-18}$Wm$^{-2}$) &
  $\bar{\nu}_{obs}$ (cm$^{-1}$)  & F($\times$10$^{-18}$Wm$^{-2}$) &
  $\nu_0$ (cm$^{-1}$) & $\nu_0$ (GHz) & $\lambda_0$ ($\mu$m)  &
  Species & Transition  \\
  \hline
    21.297 &   34.1 $\pm$   4.3 & 21.302 $\pm$ 0.005 &   47.4 $\pm$   3.8 &  ?      &           &  ?      & ?             & ?                                                 \\
    21.428 &  144.5 $\pm$  11.5 & 21.419 $\pm$ 0.002 &  117.8 $\pm$   3.8 &  21.423 &    642.25 &  466.72 & SO$_2$        & 36$_{7\,29}$-36$_{6\,30}$                         \\
    21.538 &  122.0 $\pm$  10.5 & 21.530 $\pm$ 0.002 &  110.0 $\pm$   3.8 &  21.540 &    645.75 &  464.24 & p-H$_2$O      & 9$_{73}$-8$_{82}$                                 \\
           &                    &                    &                    & +21.545 &    645.90 &  464.14 & o-H$_2$O?     & 9$_{72}$-8$_{81}$                                 \\
           &                    &                    &                    & +21.544 &    645.87 &  464.16 & SO            & 15$_{16}$-14$_{15}$                               \\
    21.717 &  222.6 $\pm$   3.7 & 21.704 $\pm$ 0.003 &  165.3 $\pm$  13.3 &  21.710 &    650.85 &  460.62 & SiO           & 15-14                                             \\
\hline
  \end{tabular}
  }
  \end{table*}
  \addtocounter{table}{-1}
  \begin{table*}
  {\footnotesize
  \centering
  \caption{-- continued.}
  \begin{tabular}{lr lr rrl ll}
  \hline
  \multicolumn{2}{l}{OD123} & \multicolumn{2}{l}{OD317} & \multicolumn{4}{l}{ID}\\\
  $\bar{\nu}_{obs}$ (cm$^{-1}$)  & F($\times$10$^{-18}$Wm$^{-2}$) &
  $\bar{\nu}_{obs}$ (cm$^{-1}$)  & F ($\times$10$^{-18}$Wm$^{-2}$) &
  $\nu_0$ (cm$^{-1}$) & $\nu_0$ (GHz) & $\lambda_0$ ($\mu$m)  &
  Species & Transition  \\
  \hline
    28.556 &  214.8 $\pm$   8.9 & 28.565 $\pm$ 0.002 &  116.0 $\pm$   4.1 &  28.555 &    856.06 &  350.20 & SiS?          & v=3, 48-47                                        \\
           &                    &                    &                    & +28.572 &    856.57 &  350.00 & Si$^{33}$S?   & 48-47                                             \\
           &                    &                    &                    & +28.575 &    856.66 &  349.95 & $^{29}$SiO?   & 20-19                                             \\
    28.692 &  211.0 $\pm$  13.0 & 28.687 $\pm$ 0.001 &  177.7 $\pm$   2.5 &  28.685 &    859.95 &  348.61 & p-H$_2$O      & $\nu_2$=1, 2$_{11}$-2$_{02}$                      \\
    28.802 &  120.7 $\pm$  12.1 & 28.797 $\pm$ 0.002 &   76.4 $\pm$   2.4 &  28.815 &    863.85 &  347.05 & o-H$_2$O?     & 10$_{83}$-9$_{90}$                                \\
           &                    &                    &                    & +28.815 &    863.85 &  347.04 & p-H$_2$O?     & 10$_{82}$-9$_{91}$                                \\
\hline
  \end{tabular}
  }
  \end{table*}
  
    \begin{table*}
  {\footnotesize
  \centering
  \caption{VY~CMa: Line fluxes and identifications in the SPIRE FTS SSW spectra
  \label{linelist-2}}
  \begin{tabular}{lr lr rrl ll}
  \hline
  \multicolumn{2}{l}{OD123} & \multicolumn{2}{l}{OD317} & \multicolumn{4}{l}{ID}\\\
  $\bar{\nu}_{obs}$ (cm$^{-1}$)  & F($\times$10$^{-18}$Wm$^{-2}$) &
  $\bar{\nu}_{obs}$ (cm$^{-1}$)  & F($\times$10$^{-18}$Wm$^{-2}$) &
  $\nu_0$ (cm$^{-1}$) &   $\nu_0$ (GHz) & $\lambda_0$ ($\mu$m)  &
  Species & Transition  \\
  \hline
    32.624 &   35.5 $\pm$   6.2 & 32.624 $\pm$ 0.009 &   45.2 $\pm$  10.1 &  32.637 &    978.43 & 306.40 & o-H$_2$O?     & $\nu_3$=2, 7$_{52}$-6$_{61}$                      \\
           &                    &                    &                    & +32.641 &    978.55 & 306.37 & CS?           & 20-19                                             \\
    32.698 &   87.7 $\pm$   6.1 & 32.690 $\pm$ 0.006 &   79.6 $\pm$  10.0 &  32.688 &    979.96 & 305.92 & SiS?          & v=3,55-54                                         \\
           &                    &                    &                    & +32.708 &    980.56 & 305.74 & Si$^{33}$S?   & 55-54                                             \\
    32.832 &   37.8 $\pm$   5.4 &                    &                    &  32.850 &    984.82 & 304.41 & $^{29}$SiO?   & 23-22                                             \\
           &                    &                    &                    & +32.851 &    984.85 & 304.40 & Si$^{34}$S?   & 56-55                                             \\
\hline
  \end{tabular}
  }
  \end{table*}
  \addtocounter{table}{-1}
  \begin{table*}
  {\footnotesize
  \centering
  \caption{-- continued.}
  \begin{tabular}{lr lr rrl ll}
  \hline
  \multicolumn{2}{l}{OD123} & \multicolumn{2}{l}{OD317} & \multicolumn{4}{l}{ID}\\\
  $\bar{\nu}_{obs}$ (cm$^{-1}$)  & F($\times$10$^{-18}$Wm$^{-2}$) &
  $\bar{\nu}_{obs}$ (cm$^{-1}$)  & F($\times$10$^{-18}$Wm$^{-2}$) &
  $\nu_0$ (cm$^{-1}$) & $\nu_0$ (GHz) & $\lambda_0$ ($\mu$m)  &
  Species & Transition  \\
  \hline
    37.339 &   97.5 $\pm$  16.9 & 37.328 $\pm$ 0.007 &   88.4 $\pm$  10.8 &  37.322 &   1118.89 & 267.94 & o-H$_2$O?     & $\nu_2$=2, 7$_{34}$-6$_{43}$                      \\
           &                    &                    &                    & +37.328 &   1119.07 & 267.89 & SiO?          & v=1,26-25                                         \\
           &                    &                    &                    & +37.356 &   1119.90 & 267.69 & SiS?          & 62-61                                             \\
    37.401 &   53.5 $\pm$  17.7 & 37.393 $\pm$ 0.010 &   47.4 $\pm$  11.1 &  37.388 &   1120.86 & 267.47 & o-H$_2$O?     & $\nu_2$=2,7$_{70}$-8$_{63}$                       \\
           &                    &                    &                    & +37.395 &   1121.07 & 267.42 & SiS?          & v=3, 63-62                                        \\
    37.591 &  308.4 $\pm$  10.6 & 37.583 $\pm$ 0.001 &  273.7 $\pm$   7.3 &  37.590 &   1126.92 & 266.03 & SiO           & 26-25                                             \\
\hline
  \end{tabular}
  }
  \end{table*}
  \addtocounter{table}{-1}
  \begin{table*}
  {\footnotesize
  \centering
  \caption{-- continued.}
  \begin{tabular}{lr lr rrl ll}
  \hline
  \multicolumn{2}{l}{OD123} & \multicolumn{2}{l}{OD317} & \multicolumn{4}{l}{ID}\\\
  $\bar{\nu}_{obs}$ (cm$^{-1}$)  & F($\times$10$^{-18}$Wm$^{-2}$) &
  $\bar{\nu}_{obs}$ (cm$^{-1}$)  & F($\times$10$^{-18}$Wm$^{-2}$) &
  $\nu_0$ (cm$^{-1}$) & $\nu_0$ (GHz) & $\lambda_0$ ($\mu$m)  &
  Species & Transition  \\
  \hline
    42.403 &   99.8 $\pm$  19.4 & 42.399 $\pm$ 0.004 &   91.1 $\pm$   9.2 &  42.394 &   1270.94 & 235.88 & CS            & 26-25                                             \\
           &                    &                    &                    & +42.402 &   1271.18 & 235.84 & o-H$_2$S?     & 7$_{34}$-7$_{25}$                                 \\
           &                    &                    &                    & +42.412 &   1271.48 & 235.78 & p-H$_2$O?     & 13$_{3\,11}$-12$_{48}$                            \\
    42.506 &   70.9 $\pm$  19.4 & 42.501 $\pm$ 0.007 &   53.4 $\pm$   9.3 &  42.502 &   1274.18 & 235.28 & SiS?          & v=1,71-70                                         \\
           &                    &                    &                    & +42.514 &   1274.54 & 235.21 & $^{29}$SiS?   & 72-71                                             \\
    42.642 &  587.2 $\pm$  21.0 & 42.635 $\pm$ 0.001 &  570.3 $\pm$  10.2 &  42.638 &   1278.26 & 234.53 & o-H$_2$O      & 7$_{43}$-6$_{52}$                                 \\
\hline
  \end{tabular}
  }
  \end{table*}
  \addtocounter{table}{-1}
  \begin{table*}
  {\footnotesize
  \centering
  \caption{-- continued.}

  \begin{tabular}{lr lr rrl ll}
  \hline
  \multicolumn{2}{l}{OD123} & \multicolumn{2}{l}{OD317} & \multicolumn{4}{l}{ID}\\\
  $\bar{\nu}_{obs}$ (cm$^{-1}$)  & F($\times$10$^{-18}$Wm$^{-2}$) &
  $\bar{\nu}_{obs}$ (cm$^{-1}$)  & F($\times$10$^{-18}$Wm$^{-2}$) &
  $\nu_0$ (cm$^{-1}$) & $\nu_0$ (GHz) & $\lambda_0$ ($\mu$m)  &
  Species & Transition  \\
  \hline
    45.767 &  183.3 $\pm$  16.4 & 45.763 $\pm$ 0.002 &  181.8 $\pm$   5.7 &  45.778 &  1372.39 & 218.45 & $^{30}$SiS?   & 79-78                                             \\
    45.891 &  151.6 $\pm$  16.4 & 45.912 $\pm$ 0.001 &  506.5 $\pm$   6.0 &  45.900 &  1376.05 & 217.86 & SiO           & v=1, 32-31                                        \\
           &                    &                    &                    & +45.893 &  1375.84 & 217.90 & o-H$_2^{18}$O & $\nu_2$=1, $3_{21}$-$3_{12}$                      \\
    46.098 & 1296.0 $\pm$  17.6 & 46.098 $\pm$ 0.001 & 1276.0 $\pm$   6.4 &  46.098 &  1381.98 & 216.93 & CO            & 12-11                                             \\
    46.210 &  319.6 $\pm$  16.5 & 46.215 $\pm$ 0.001 &  324.5 $\pm$   5.6 &  46.220 &  1385.64 & 216.35 & SiO?          & 32-31                                             \\
    46.380 &  397.9 $\pm$  16.5 & 46.375 $\pm$ 0.001 &  443.1 $\pm$   5.8 &  46.362 &  1389.90 & 215.69 & SO$_2$?       & 17$_{11\,7}$-16$_{10\,6}$                         \\
\hline
  \end{tabular}
  }
  \end{table*}
  \addtocounter{table}{-1}
  \begin{table*}
  { \footnotesize
  \centering
  \caption{-- continued.}

  \begin{tabular}{lr lr rrl ll}
  \hline
  \multicolumn{2}{l}{OD123} & \multicolumn{2}{l}{OD317} & \multicolumn{4}{l}{ID}\\\
  $\bar{\nu}_{obs}$ (cm$^{-1}$)  & F($\times$10$^{-18}$Wm$^{-2}$) &
  $\bar{\nu}_{obs}$ (cm$^{-1}$)  & F($\times$10$^{-18}$Wm$^{-2}$) &
  $\nu_0$ (cm$^{-1}$) & $\nu_0$ (GHz) & $\lambda_0$ ($\mu$m)  &
  Species & Transition  \\
  \hline
    49.473 &   57.5 $\pm$   7.9 & 49.448 $\pm$ 0.004 &   44.2 $\pm$   4.1 &  49.477 &  1483.28 & 202.11 & CO?           & v=1, 13-12                                        \\
           &                    &                    &                    & +49.479 &  1483.34 & 202.10 & $^{29}$SiS?   & 84-83                                             \\
           &   87.7 $\pm$   7.8 &                    &   87.7 $\pm$   7.8 &  49.563 &  1485.86 & 201.76 & SiS?          & v=1, 83-82                                        \\
    49.656 &   99.0 $\pm$   6.6 & 49.652 $\pm$ 0.002 &  101.7 $\pm$   4.1 &  49.626 &  1487.75 & 210.51 & p-H$_2$O?     & $\nu_2$=1,$\nu_3$=1, 8$_{45}$-7$_{52}$            \\
           &                    &                    &                    & +49.653 &  1488.56 & 201.40 & SiS?          & v=3, 84-83                                        \\
           &                    &                    &                    & +49.682 &  1489.43 & 201.28 & o-H$_2$O?     & $\nu_3$=1, 9$_{64}$-8$_{71}$                      \\
\hline
   \end{tabular}
   }  
   \end{table*}
  

\newpage

\label{lastpage}

\end{document}